\documentclass[twocolumn]{aastex702}

\usepackage[flushleft]{threeparttable}
\usepackage{array,booktabs,makecell}
\usepackage{subfigure}
\usepackage{orcidlink}
\usepackage{tikz}
\usepackage{multirow}
\usepackage{hyperref}
\usepackage{graphicx}
\usepackage{natbib}
\usepackage{placeins}
\usepackage{amsmath}
\usepackage{tgcursor}
\usepackage{rotating}
\usepackage{tablefootnote}
\usepackage{tabularx}
\def\arraybackslash{\let\\\tabularnewline} 

\newcommand{\hcop}{HCO$^+$}
\newcommand{\dcop}{DCO$^+$}
\newcommand{\htcop}{H$^{13}$CO$^+$}

\newcommand{\nthp}{N$_2$H$^+$}
\newcommand{\ntdp}{N$_2$D$^+$}
\newcommand{\htdp}{H$_2$D$^+$}
\newcommand{\htp}{H$_{3}^+$}
\newcommand{\hcoh}{HCO$^+$ $4 - 3$}
\newcommand{\htcol}{H$^{13}$CO$^+$ $3 - 2$}
\newcommand{\nthl}{N$_2$H$^+$ $3 - 2$}
\newcommand{\nthh}{N$_2$H$^+$ $4 - 3$}
\newcommand{\dcol}{DCO$^+$ $4 - 3$}
\newcommand{\dcoh}{DCO$^+$ $5 - 4$}
\newcommand{\solarmass}{M$_{\odot}$}
\def\farcs{\hbox{$.\!\!^{\prime\prime}$}} 

\shorttitle{Disk Ionization Survey}
\shortauthors{Long et al.}

\graphicspath{{./}{figures/}}

\begin{document}

\title{Mapping Molecular Ion Distributions in a Diverse Sample of Protoplanetary Disks}

\author{Deryl E. Long \orcidlink{0000-0003-3840-7490}} \affiliation{Department of Astronomy, University of Virginia, Charlottesville, VA 22904, USA}
\email{del6h@virginia.edu}

\author{L. Ilsedore Cleeves \orcidlink{0000-0003-2076-8001}}
\affiliation{Department of Astronomy, University of Virginia, Charlottesville, VA 22904, USA}
\email{lic3f@virginia.edu}

\author{Charles J.\ Law \orcidlink{0000-0003-1413-1776}}
\affiliation{Department of Astronomy, University of Virginia, Charlottesville, VA 22904, USA}
\affiliation{Minnesota Institute for Astrophysics, University of Minnesota, 116 Church St. SE, Minneapolis, MN 55455}
\email{cjl8rd@virginia.edu}  

\author{Fred C. Adams \orcidlink{0000-0002-8167-1767}}
\affiliation{Physics Department, University of Michigan, Ann Arbor, MI 48109, USA}
\email{fca@umich.edu}

\author{Dana E. Anderson \orcidlink{0000-0002-8310-0554}}
\affiliation{Volgenau Department of Physics, U.S. Naval Academy, Annapolis, MD, USA}
\email{danderso@usna.edu}

\author{Sean M. Andrews \orcidlink{0000-0003-2253-2270}}
\affiliation{Harvard-Smithsonian Center for Astrophysics, 60 Garden St., Cambridge, MA 02138, USA}
\email{sandrews@cfa.harvard.edu}

\author{Edwin A. Bergin \orcidlink{0000-0003-4179-6394}}
\affiliation{Department of Astronomy, University of Michigan, 1085 South University Avenue, Ann Arbor, MI 48109, USA}
\email{ebergin@umich.edu}

\author{Jane Huang \orcidlink{0000-0001-6947-6072}}
\affiliation{Columbia University, 538 West 120th Street, New York, NY 10027, USA}
\email{jane.huang@columbia.edu}

\author{A. Meredith Hughes \orcidlink{0000-0002-4803-6200}}
\affiliation{Van Vleck Observatory, Wesleyan University, 96 Foss Hill Dr., Middletown, CT 06459, USA}
\email{amhughes@wesleyan.edu}

\author{Chunhua Qi \orcidlink{0000-0001-8642-1786}}
\affiliation{Institute for Astrophysical Research, Boston University, 725 Commonwealth Avenue, Boston, MA 02215, USA}
\email{cqi1@bu.edu}

\author{Kamber R. Schwarz \orcidlink{0000-0002-6429-9457}}
\affiliation{Department of Physics and Astronomy, University of Georgia, 220 Cedar St Physics Building, Athens, GA 30602}
\email{kamberschwarz@uga.edu}

\author{Jacob B. Simon \orcidlink{0000-0002-3771-8054}}
\affiliation{Department of Physics and Astronomy, Iowa State University, Ames, IA, 50010, USA}
\email{jbsimon@iastate.edu}

\author{David Wilner \orcidlink{0000-0003-1526-7587}}
\affiliation{Harvard-Smithsonian Center for Astrophysics, 60 Garden St., Cambridge, MA 02138, USA}
\email{dwilner@cfa.harvard.edu}

\begin{abstract}

Observations of ionization-tracing molecules in protoplanetary disks suggest that their ionization environments are diverse due to varying degrees of cosmic ray modulation, stellar activity, and environmental differences that alter the impact of ionizing radiation. To better understand ionization chemistry in disks, we present the largest survey of resolved observations of molecular ions to date. Using new and archival ALMA observations we detect \hcop, \htcop, \nthp, and \dcop \ toward seven protoplanetary disks (AS 209, DM Tau, GM Aur, HD 163296, LkCa 15, MWC 480, and V4046 Sgr). We compare distributions of ions to continuum features and find no robust relationships between the two. The observations do, however, reveal distinctions between Herbigs and T-Tauris and hint at a dichotomy between two types of T-Tauri ionization environments. The Herbig sources exhibit unique centrally-peaked \nthp\ $J = 4-3$ emission, which we speculate could be tracing either a highly ionized surface layer or a zone of CO destruction, possibly due to their strong UV fields. Three T-Tauri sources in our sample exhibit bright, radially coincident rings in all optically thin molecular ion lines. We speculate that these radially coincident ion rings are evidence of a global change in ionization, possibly at the edge of a T-Tauriosphere. While it is not yet clear what factors most strongly influence the distribution of ion emission, this survey more than doubles the number of resolved ion observations in disks and points to potential distinctions in ion emission morphology related to the central star and environment. 

\end{abstract}


\section{Introduction} \label{sec:intro}

The degree of ionization in a protoplanetary disk determines both its chemical and physical evolution. Ionization is especially important in the cold, dense midplane where planet formation occurs. Ions fuel gas-phase chemistry at low temperatures \citep{1973ApJ...185..505H}, creating important chemical pathways for molecule formation, including water \citep{vandishoeck2013} and key organic species \citep{cleeves16}. On icy grain mantles, hydrogenation reactions are powered by hydrogen atoms produced via ionization of molecular H$_{2}$ \citep{hasegawa_92}. 

\begin{deluxetable*}{lclllllllll} 
	\tabletypesize{\footnotesize}
	\tablecaption{Source properties. \label{tab:source_props}}
	\tablecolumns{11} 
	\tablewidth{\textwidth} 
	\tablehead{
		\colhead{Source}          &
		\colhead{Dist.$^{a}$}       &
		\colhead{Age}             & 
		\colhead{$M_\star$}       & 
		\colhead{$L_\star$}       & 
        \colhead{L$_{XR}$$^{b}$}         &
            \colhead{log$_{10}$($\dot{M}$)} &
		\colhead{$v_{sys}$}       &
		\colhead{Inc.}            & 
		\colhead{PA}              &
            \colhead{Source flags$^{c}$}     \\
            \colhead{}               &
		\colhead{[pc]}            &
		\colhead{[Myr]}           & 
		\colhead{[$M_\odot$]}     & 
		\colhead{[$L_\odot$]}     &
        \colhead{[$L_\odot$]}       &
            \colhead{[$M_\odot$\ yr$^{-1}$]} &
		\colhead{[km s$^{-1}$]}   &
		\colhead{[deg]}           & 
		\colhead{[deg]}           &
            \colhead{}
		  }
\startdata
AS 209 & 121.0 & 1.6$^{[1]}$ & 1.2$^{[8]}$ & 1.5$^{[1]}$ & 2.5$\times 10^{-4}$ & -7.5$^{[18]}$ & 4.6 $^{[8]}$ &  35.0$^{[22]}$& 85.8$^{[22]}$ & TT \\
DM Tau & 145.1 & 3.5 -- 8$^{[2, 3]}$ & 0.45$^{[16]}$ & 0.2$^{[3]}$& 7.0$\times 10^{-5}$ & -8.5$^{[13]}$ & 6.0 $^{[10]}$ & 36.0 $^{[12]}$ & 155.6$^{[12]}$  & TT, TD  \\
GM Aur & 159.0 & 2.1$^{[3]}$ & 1.1$^{[8]}$ & 1.2$^{[3]}$ & 1.7$\times 10^{-4}$ & -8.1$^{[19]}$ & 5.6 $^{[8]}$ & 53.2$^{[25]}$ & 57.2$^{[25]}$ & TT, TD \\
HD 163296 & 101.5 & 7.6$^{[5]}$ & 2.0 $^{[8]}$ & 15.8 $^{[5]}$& 1.0$\times 10^{-4}$ $^{[23]}$ & -7.4$^{[20]}$ & 5.8 $^{[8]}$ & 46.7$^{[22]}$ &  133.3$^{[22]}$ & HAe \\
LkCa 15 & 158.9 & 3 -- 5$^{[2, 3, 4]}$ & 1.17$^{[16]}$ & 0.8$^{[3]}$& 2.5$\times 10^{-4}$ & -9.2$^{[14]}$ & 6.3 $^{[10]}$ & 50.6$^{[12]}$ & 61.6$^{[12]}$ & TT, TD  \\
MWC 480 & 161.8 & 6.2$^{[5]}$ & 2.1$^{[8]}$ & 18.6$^{[5]}$& 1.0$\times10^{-4}$ & -6.9$^{[21]}$ & 5.1$^{[8]}$  & 37.0$^{[26]}$ & 148.0$^{[26]}$ & HAe \\
V4046 Sgr & 72.4 & 5 -- 18.5$^{[6, 17]}$& 1.76$^{[16], d}$ & 0.49, 0.33$^{[7]}$ & 3.1$\times 10^{-4}$ $^{[24]}$ & -9.3$^{[15]}$ & 2.8$^{[11]}$ & 33.4$^{[12]}$ & 76.0$^{[12]}$ & TD, CB\\
\enddata
\tablenotetext{a}{From \textit{Gaia} Data Release 2 \citep{Gaia2018}}
\tablenotetext{b}{When published L$_{XR}$ values are not available we calculate L$_{XR}$ using stellar X-ray fluxes and/or count rates from the High Energy Astrophysics Science Archive Research Center (HEASARC).}
\tablenotetext{c}{Source property flags: T-Tauri (TT), transition disk (TD), Herbig Ae (HAe), circumbinary (CB).}
\tablenotetext{d}{Combined dynamical mass for both components of the spectroscopic binary \citep{izquierdo_25}.}
\tablecomments{References: [1] \citet{andrews_2009}, [2] \citet{Guilloteau2014}, [3] \citet{andrews_2013},  [4] \citet{Simon_2000}, [5] \citet{vioque_2018}, [6] \citet{rosenfeld_2012}, 
[7] \citet{quast_2000}, [8] \citet{teague_21}, [9] \citet{oberg_maps_2021}, [10] \citet{Law_2023}, [11] \citet{Kastner_2018}, [12] \citet{curone_2025}, [13] \citet{france14}, [14] \citet{donati19}, [15] \citet{donati11}, [16] \citet{izquierdo_25}, [17] \citet{miret-rouig_2020}, [18] \citet{salyk_13}, [19] \citet{ingleby_2015}, [20] \citet{fairlamb_15}, [21] \citet{mendigutia_13}, [22] \citet{Huang_2018}, [23] \citet{swartz_05}, [24] \citet{sacco_12}, [25] \citet{Huang_2020}, [26] \citet{liu_2019}}
\end{deluxetable*}

In addition to driving chemistry and influencing the composition of forming planets, ionization also impacts the distribution of disk material and the gas dynamics, which are governed by hydro-- and magneto--hydrodynamic (MHD) processes. When sufficiently ionized, gas can couple to magnetic field lines and drive a disk wind \citep{blandford_1982} and/or the magneto-rotational instability \citep[MRI;][]{balbus91}. In the most well-ionized regions of the disk, we expect turbulence to be driven primarily by the MRI. In weakly-ionized regions, non-ideal effects such as Ohmic diffusion \citep{flemming03,gole16}, ambipolar diffusion \citep{simon13a,simon13b}, and the Hall effect \citep{bai15,simon15b} alter the strength and degree of isotropy of background turbulence, if present. Hydrodynamic instabilities such as the ``vertical shear instability" \citep[VSI;][]{nelson_2013}, ``convective overstability" \citep[COS;][]{klahr_2014}, and the ``zombie vortex instability" \citep[ZVI;][]{marcus_2015} become important in disk regions where ionization is extremely low. 

Ionization and its relationship to disk turbulence are particularly critical during the process of planetesimal formation. Planetesimals on scales of $\sim$0.1--100 km are formed from micron-sized dust grains. To grow from microns to km in size, these objects must overcome radial drift \citep{whipple_1972} and fragmentation \citep{blum_2018}. The "streaming instability" (SI) may help to overcome these barriers by promoting planetesimal formation via gravitational collapse \citep{youdin_2005,johansen07,simon_2016}. However, turbulence may suppress the SI (\citealt{Chen_2020}, \citealt{umurhan20}, \citealt{gole20}, \citealt{lim23}, but see e.g., \citealt{johansen07}, \citealt{xu_bai_22}). The degree of ionization has significant influence on the gas dynamics of the disk, which in turn impact the processes governing planet formation. 

 \begin{figure*}
    \centering
    \includegraphics[scale=0.3]{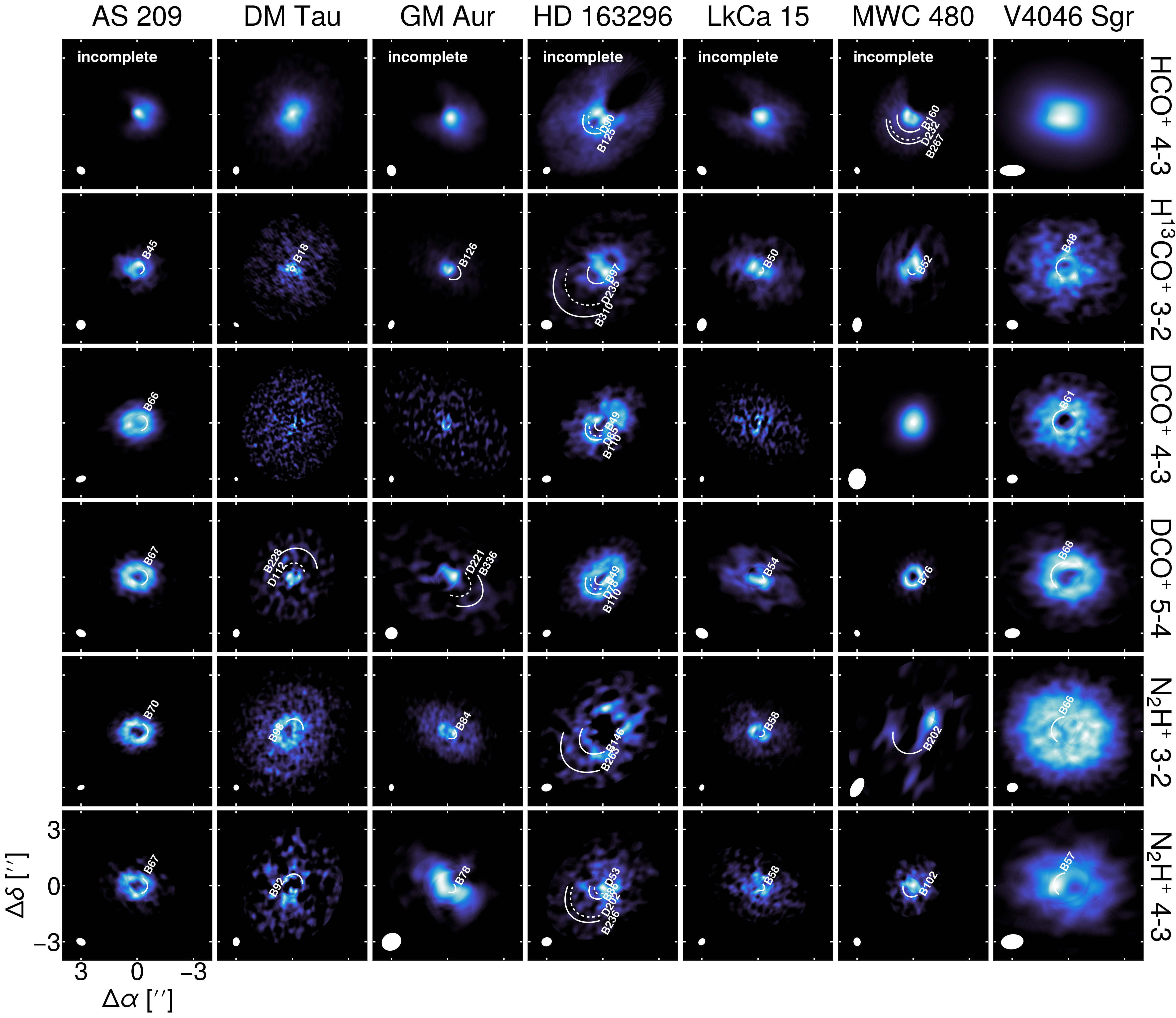}
        \caption{\normalsize{Zeroth moment maps of the target molecular ions across the sample. Rings and gaps from Table \ref{tab:ion_substr} are labeled and shown as solid and dotted arcs, respectively. Beam sizes are shown in the lower left corner of each panel. We note the data cut-off issue (i.e., insufficient spectral range that does not cover the full line profile) which effects the \hcop\ cubes for AS 209, GM Aur, HD 163296, LkCa 15, and MWC 480 (see Section \ref{sec:data_obs}).}}
  \label{fig:mom0}
\end{figure*}

The main ionizing sources in disks are ultraviolet (UV) photons, X-rays, and cosmic rays (CRs). The dominant source varies throughout the disk and is set by the disk density and penetrating power of the radiation. UV photons are a powerful source of ionization at the disk surface but are attenuated at depths greater than 0.01--0.1 g cm$^{-2}$ \citep{perez-becker11b}. X-rays penetrate further into the disk, reaching depths up to $\sim$10 g cm$^{-2}$ depending on the hardness of the stellar spectrum \citep{igea}. CRs have the greatest penetrating power, ionizing gas column densities up to $\sim$100 g cm$^{-2}$ and thus, are expected to be the dominant ionization source in the disk midplane. However, the CR ionization rate ($\zeta_{CR}$) could be reduced by orders of magnitude due to CR exclusion \citep{cleeves13a,cleeves15_twhya} or enhanced by local particle acceleration \citep{padovani_2015, padovani_2016, brunn_24}. 

Stellar photoionization in the form of UV photons contributes primarily to the ionization in low-density surface layers, producing singly-ionized atomic species such as C$^{+}$, O$^{+}$, and N$^{+}$. XRs and CRs penetrating the dense gas will ionize molecular H$_{2}$, leading to the production of H$_{3}^{+}$. H$_{3}^{+}$ goes on to drive ion chemistry in disks, creating formation pathways for other molecular ions including \hcop, \nthp, and their deuterated isotopologues \citep{bergin07,Willacy_2007,Aikawa_2015,Aikawa_2018}. The bulk of the ionized gas in \htp\ is not observable with existing facilities, and the observable deuterated isotopologue \htdp\ has yet to be detected in a protoplanetary disk. However, \hcop, \nthp, \ntdp, and \dcop\ have all been detected in disks. In fact, molecular ions have been targeted in several (sub)millimeter surveys with ALMA and the SMA. An ALMA survey of deuterium chemistry yielded  the first detection of \ntdp\ in a protoplanetary disk \citep{huang15} as well as detections of \dcop\ and \htcop\ in six disks \citep{huang17}. A snowline survey with ALMA yielded detections of \nthp\ in six disks and revealed varying \nthp\ emission morphologies \citep{qi_2019}. The MAPS survey detected \hcop\ and \htcop\ in their five target disks and found that midplane ionization rates appear to vary across the sample \citep{Aikawa_2021}. Spatially-resolved observations of multiple molecular ions have yielded increasingly detailed constraints on the distribution and driver(s) of ionization in TW Hya, IM Lup, and DM Tau \citep{cleeves15_twhya, Teague_2015, Seifert_2021, long24}. 

In this work, we present the most comprehensive observational multi-line ionization study to date in a diverse sample of protoplanetary disks. We examine in detail new and archival high-resolution observations of multiple ions expected to trace different emitting layers spanning from the warm molecular layer to the cold midplane. In Section \ref{sec:sample} we give an overview of the source properties for the star-disk systems observed in this survey. In Section \ref{sec:obs}, we discuss the observational details and data analysis. In Section \ref{sec:results}, we present the observational results, highlighting the radial distributions of molecular ions in our sample. In Section \ref{sec:disc}, we discuss the observed relationships between the distribution of molecular ions and physical properties as well as implications for disk chemistry. Ultimately, we leverage these sensitive data to better understand ion chemistry in disks and pave the way for more accurate and informative interpretations of molecular ion observations going forward. 

\section{Sample}\label{sec:sample}
\begin{deluxetable*}{lcllc} 
	\tabletypesize{\footnotesize}
	\tablecaption{Summary of ALMA observations. \label{tab:obs}}
	\tablecolumns{5} 
	\tablewidth{\textwidth} 
	\tablehead{
		\colhead{Project Code}          &
            \colhead{PI}                    &
		\colhead{Target Molecule}      &
		\colhead{Source}             & 
		\colhead{Project ID}       
		  }
\startdata
2012.1.00681.S & C. Qi & \nthl, \dcol & HD 163296 & P2012$^{[a]}$ \\ \hline
2013.1.00226.S & K. \"Oberg & \htcol & AS 209, HD 163296, LkCa 15,  &  P2013 \\
               &       &        &  MWC 480, V4046 Sgr  &   \\ \hline
2015.1.00657.S & K. \"Oberg & \nthl, \dcol & MWC 480 & P2015-1$^{[b]}$ \\ \hline
2015.1.00678.S & C. Qi & \nthl, \dcol & AS 209, DM Tau, GM Aur, &  P2015-2$^{[c]}$ \\
               &    &              & HD 163296, LkCa 15, V4046 Sgr  \\ \hline
2016.1.00956.S & V. Guzmán & \hcoh  & V4046 Sgr & P2016   \\ \hline
2019.1.00379.S & L.I. Cleeves & \hcoh, \htcol, & DM Tau, GM Aur &  P2019   \\
               &         & \nthh, \dcoh   &    &        \\ \hline
2021.1.00138.S & L.I. Cleeves & \hcoh, \htcol, &  AS 209, GM Aur, HD 163296,  & P2021  \\ 
               &         & \nthh, \dcoh   &  LkCa 15, MWC 480, V4046 Sgr  &         \\ 
\enddata
\tablenotetext{a}{Data first published in \citet{qi_2015}.}
\tablenotetext{b}{Data first published in \citet{loomis_2020}.}
\tablenotetext{c}{Data first published in \citet{qi_2019}.}
\end{deluxetable*}

Table \ref{tab:source_props} summarizes source details for the 7 protoplanetary disks in our sample. Our target disks were selected to span a range of ages ($\leq$ 2 Myr -- $\sim$20 Myr) and the sample exhibits diverse disk morphologies. We target three T-Tauri disks with clear inner-disk dust gaps, DM Tau, GM Aur, and LkCa 15; one T Tauri disk without a significant inner-disk dust gap, AS 209; two Herbig Ae disks, HD 163296 and MWC 480; and one circumbinary disk, V4046 Sgr. The sources also vary in stellar mass and X-ray luminosity, allowing us to explore how stellar and disk properties impact ionization chemistry. 

 \begin{figure*}[t!]
    \centering
    \includegraphics[scale=0.26]{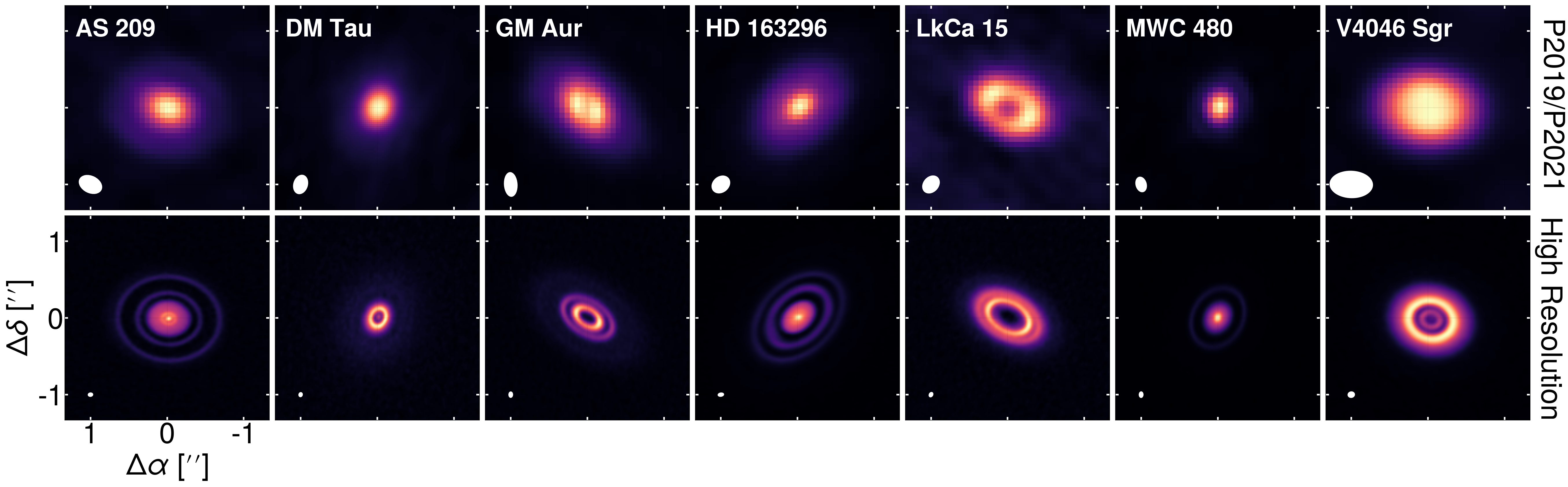}
        \caption{\normalsize{Top row: Band 7 continuum images from Projects 2019.1.00379.S and 2021.1.00138.S. Corresponding flux and beam information can be found in Table \ref{tab:cont}. Bottom row: High-resolution continuum images from MAPS and exoALMA at 260 GHz and 331 GHz, respectively \citep{law_2021_rad,Sierra21,curone_2025}.}}
  \label{fig:cont}
\end{figure*}

AS 209 is a 1.2 \solarmass\ T-Tauri star \citep{andrews_2009, teague_21} with a $\sim$180 au millimeter dust disk and gas emission extending out to $\sim$270 au \citep{Huang_2018,guzman2018}. The millimeter dust disk has numerous substructures, including seven gap/ring pairs identified by \citet{Huang_2018}. This disk also hosts a circumplanetary disk (CPD) candidate seen in $^{13}$CO gas located at a radius of $\approx$ 200 au \citep{bae_22}.

DM Tau is a low mass \citep[0.45 \solarmass;][]{izquierdo_25} T-Tauri star that hosts a large transition disk. High resolution millimeter observations of DM Tau reveal a dust cavity bounded by inner and outer rings at radii of $\approx$ 3 and 20 au \citep{Kudo_2018}. DM Tau is one of only two disks with evidence of gas turbulence \citep{flaherty, flaherty24}.

GM Aur is a 1.1 \solarmass\ T-Tauri star \citep{mcclure2016} that hosts a transition disk with a $\sim$40 au cavity \citep{macias18}. Interferometric observations of CO revealed non-Keplerian signatures in the inner disk indicative of a warp \citep{hughes_2009}, possibly induced by a planet. GM Aur is thus a popular candidate for studying ``active" gap clearing by a planet. \citet{huang_21} also identified multiple large-scale gas structures in the GM Aur system, including four spiral arms, which they propose to be evidence of late infall from the protostellar envelope or surrounding cloud material.

HD 163296 (also known as MWC 275) is an intermediate-mass Herbig AeBe star \citep[2.0 \solarmass;][]{vioque_2018} with an active bi-polar jet \citep[HH409;][]{grady_2000, devine_2000}. HD 163296 exhibits dust substructure \citep{Isella_2018} and is host to three $\sim$M$_{J}$ planet candidates, identified from disk kinematics, at 83, 137, and 260 au \citep{Teague_2018,pinte_2018}.

LkCa 15 is a $\sim$1.2 \solarmass \ T-Tauri star \citep{Simon_2000,andrews_2013} that hosts a transition disk with a 76 au dust cavity \citep{francis20}. High-resolution ALMA Band 6 observations revealed additional narrow substructures in the inner disk and millimeter emission indicating the presence of small grains within the dust cavity \citep{facchini20}. \citet{leemker_2022} found evidence of a $\sim$10 au gas cavity in LkCa 15, however the amount of gas still present within the dust cavity rule outs the possibility of gas clearing by massive planets. Instead, it is possible that there are several lower-mass planets present within the cavity in LkCa 15. 

MWC 480 (also known as HD 31648) is an intermediate-mass Herbig AeBe star \citep[2.1 \solarmass;][]{vioque_2018}. The millimeter dust disk has an effective radius of 105 au \citep{Long_2018} and two gap/ring pairs \citep{liu_2019, Sierra21}. Hydrodynamic simulations suggest that the amount of mass depleted in the dust gap at 74 au is consistent with the presence of a $\sim$2.3 M$_{J}$ planet, while velocity perturbations in CO gas indicate the presence of a $\sim$1 M$_{J}$ planet at a radius of $\approx$ 245 au \citep{teague_21}. 

V4046 Sgr is a close binary system (P $\sim$2.4 day) with two stars of similar mass \citep[0.9 \solarmass\ and 0.85 \solarmass;][]{rosenfeld_2012} orbited by a circumbinary disk. The millimeter dust disk is a transition disk with an inner cavity and ring at $\sim$30 au \citep{rosenfeld2013}. High resolution millimeter continuum observations show additional narrow substructures within the cavity \citep{martinez-brunner_2022, curone_2025}. 

Our sources span a variety of environments from isolated systems to relatively dense star-forming environments. Four of the sources (DM Tau, GM Aur, LkCa 15, and MWC 480) are members of the Taurus star-forming region (SFR) with a median age of $\sim$ 1 Myr \citep{gomez_hartmann_kenyon_1993}. GM Aur and MWC 480 are close to one another on one side of the Taurus periphery, DM Tau is on the opposite side of the Taurus periphery, and LkCa 15 is near the region's center \citep[e.g.,][]{garufi_24}. V4046 Sgr is a member of the $\beta$ Pictoris moving group, which is made up of dozens of young stars with a median age of 18.5 Myr \citep{miret-rouig_2020}. AS 209 is a member of the Ophiuchus SFR, which might be impacted by feedback from massive stars in the Upper Scorpius region \citep{preibisch_1999}. HD 163296 is an isolated system.

Published X-ray luminosities exist for V4046 Sgr and HD 193296 from \textit{XMM-Newton} and \textit{Chandra} observations, respectively \citep{swartz_05,sacco_12}. For the remaining five sources, we queried the High Energy Astrophysics Science Archive Research Center (HEASARC) archive, prioritizing \textit{Chandra} data for its high sensitivity and arcsecond spatial resolution. When multiple measurements were available, we adopted the lowest value in order to maintain consistency in our selections and represent the baseline, quiescent X-ray state of the star. \textit{Chandra} X-ray fluxes (0.3–8 keV) were obtained for DM Tau, GM Aur, LkCa 15, and MWC 480, while an \textit{Einstein} count rate (0.1–6.4 keV) was used for AS 209. All fluxes were derived assuming a power-law model and the X-ray luminosity (L$_{XR}$) was then calculated using the distances in Table \ref{tab:source_props}. The resulting L$_{XR}$ values range from 7$\times$10$^{-5}$ $L_\odot$ for DM Tau up to 3.1$\times$10$^{-4}$ $L_\odot$ for the circumbinary system V4046 Sgr. 

\section{Observations}\label{sec:obs}

\subsection{Observational Details and Archival Data}

In this survey, we leverage both new and archival ALMA programs, summarized in Table \ref{tab:obs}. For all of the sources in our sample we observe four key molecular ions: \hcop, \htcop, \dcop, and \nthp. Two transitions are covered for both \nthp\ and \dcop. Previously unpublished Band 6 and 7 observations were taken in ALMA Cycles 7 and 8 under project codes 2019.1.00379.S and 2021.1.00138.S at high spatial resolution (0.2-0.4\farcs). The band 6 observations included a HC$^{18}$O$^{+}$ $J=3-2$ window at 255 GHz, however the line was not detected and is thus not included in our analysis. Full details for the new observations can be found in Appendix \ref{sec:app_obsdat} in Table \ref{tab:alma_details_new}.

Self-calibration was performed for the P2019 and P2021 datasets. We executed two to three rounds of phase calibration with solution intervals of \texttt{inf}, 60 seconds, and 30 seconds, respectively. The continuum RMS improved by a factor of 3 to 10 for individual measurement sets. Self-calibration was stopped either when the beam size began to increase significantly or when there ceased to be any improvement in the continuum RMS. The archival datasets (P2012, P2013, P2015-1, P2015-2, and P2016) were calibrated using the appropriate CASA pipeline version \citep{casa_mcmullin_2007,CASA_22}. 

For all datasets, we performed continuum subtraction (\texttt{fitorder} = 0) and line imaging using the \texttt{uvcontsub} and \texttt{tclean} tasks from CASA \texttt{v5.6.1}. For line imaging, we used automasking with a noise threshold of 4.25 for bright lines and a threshold of 3.25 in cases where emission was faint. \hcop\ and \dcop\ were imaged using Briggs weighting and a \texttt{robust} parameter of 0.5, while \nthp\ and \htcop\ were imaged using natural weighting (\texttt{robust} = 2) to maximize sensitivity. 

\subsection{Spectral Line Analysis}\label{sec:data_obs}

To analyze the spectral line data we first created a set of Keplerian masks for each source and molecular line combination with the \texttt{keplerian\textunderscore mask}\footnote{\url{https://github.com/richteague/keplerian_mask}} Python package \citep{teague_2020_kep} and source properties listed in Table \ref{tab:source_props}. We checked that the Keplerian masks were well fit to the data (i.e., appropriate radial extent and emission coverage) before using the masks to generate velocity-integrated intensity (zeroth moment) maps using the \texttt{bettermoments}\footnote{\url{https://github.com/richteague/bettermoments}}\ Python package \citep{teague_2018_bm} (Figure \ref{fig:mom0}). We then generated radial intensity profiles (Figure \ref{fig:radprofs}) and spectra using the \texttt{radial\textunderscore profile} and \texttt{integrated\textunderscore spectrum} functions, respectively, in the \texttt{GoFish}\footnote{\url{https://github.com/richteague/gofish}} \ Python package \citep{GoFish} to deproject the zeroth moment maps. The full gallery of spectra is shown in Figure \ref{fig:spec} Appendix \ref{app:spec}.

Line fluxes were measured within the Keplerian masks and are reported in Table \ref{tab:fluxes}. Uncertainties on the line fluxes were estimated by taking the standard deviation of the integrated flux within the same mask in sets of line-free channels. These uncertainties were then added in quadrature with the 10$\%$ calibration uncertainty for ALMA. We report the per channel RMS in line-free channels. Integration ranges were determined to be the first and last channels containing emission above 3$\times$RMS. 0.9 mm continuum fluxes were measured via Gaussian fitting within an elliptical aperture centered on the source, and errors were estimated via bootstrapping (Table \ref{tab:cont}).

\begin{table*}[]
\begin{threeparttable}
    \centering
    \caption{Line Observations.}
    \begin{tabularx}{\linewidth}{llcccccc}
    \hline\hline \noalign {\smallskip}
     Source & Transition & Frequency & Int. flux$^{(1)}$ & Channel Width & RMS & Beam (PA) \\
     &   & (GHz) & (Jy km~s$^{-1}$)  & (km s$^{-1}$) & (mJy beam$^{-1}$) &   {\smallskip}\\
    \hline
\hline
AS 209 & HCO$^{+}$ $J = 4 - 3$ & 356.734 & 8.57 $\pm$ 0.87 & 0.119 & 6.4 & 0\farcs48×0\farcs38 (55.0°) \\
 & H$^{13}$CO$^{+}$ $J = 3 - 2$ & 260.255 & 0.37 $\pm$ 0.04 & 0.119 & 6.6 & 0\farcs51×0\farcs48 (-3.0°) \\
 & N$_2$H$^{+}$ $J = 3 - 2$ & 279.512 & 0.78 $\pm$ 0.08 & 0.139 & 3.3 & 0\farcs38×0\farcs24 (-65.0°) \\
 & N$_2$H$^{+}$ $J = 4 - 3$ & 372.672 & 1.19 $\pm$ 0.14 & 0.119 & 21 & 0\farcs50×0\farcs36 (62.0°) \\
 & DCO$^{+}$ $J = 4 - 3$ & 288.144 & 0.89 $\pm$ 0.09 & 0.129 & 2.9 & 0\farcs54×0\farcs33 (-70.0°) \\
 & DCO$^{+}$ $J = 5 - 4$ & 360.170 & 0.90 $\pm$ 0.13 & 0.119 & 5.5 & 0\farcs53×0\farcs37 (63.0°) \\
\hline
DM Tau & HCO$^{+}$ $J = 4 - 3$ & 356.734 & 7.03 $\pm$ 0.71 & 0.119 & 8.4 & 0\farcs42×0\farcs32 (-6.0°) \\
 & H$^{13}$CO$^{+}$ $J = 3 - 2$ & 260.255 & 0.85 $\pm$ 0.09 & 0.149 & 3.2 & 0\farcs31×0\farcs18 (52.0°) \\
 & N$_2$H$^{+}$ $J = 3 - 2$ & 279.512 & 1.57 $\pm$ 0.17 & 0.139 & 3.0 & 0\farcs22×0\farcs18 (17.0°) \\
 & N$_2$H$^{+}$ $J = 4 - 3$ & 372.672 & 1.14 $\pm$ 0.17 & 0.119 & 36 & 0\farcs45×0\farcs35 (-3.0°) \\
 & DCO$^{+}$ $J = 4 - 3$ & 288.144 & 0.70 $\pm$ 0.08 & 0.129 & 2.5 & 0\farcs20×0\farcs16 (19.0°) \\
 & DCO$^{+}$ $J = 5 - 4$ & 360.170 & 0.38 $\pm$ 0.05 & 0.119 & 4.7 & 0\farcs43×0\farcs34 (-12.0°) \\
\hline
GM Aur & HCO$^{+}$ $J = 4 - 3$ & 356.734 & 10.86 $\pm$ 1.10 & 0.119 & 6.9 & 0\farcs58×0\farcs47 (11.0°) \\
 & H$^{13}$CO$^{+}$ $J = 3 - 2$ & 260.255 & 0.87 $\pm$ 0.09 & 0.149 & 2.7 & 0\farcs47×0\farcs29 (-20.0°) \\
 & N$_2$H$^{+}$ $J = 3 - 2$ & 279.512 & 0.93 $\pm$ 0.09 & 0.139 & 2.8 & 0\farcs36×0\farcs24 (-1.0°) \\
 & N$_2$H$^{+}$ $J = 4 - 3$ & 372.672 & 1.05 $\pm$ 0.14 & 0.119 & 41 & 1\farcs06×0\farcs90 (-61.0°) \\
 & DCO$^{+}$ $J = 4 - 3$ & 288.144 & 0.14 $\pm$ 0.02 & 0.129 & 2.4 & 0\farcs35×0\farcs22 (-2.0°) \\
 & DCO$^{+}$ $J = 5 - 4$ & 360.170 & 0.14 $\pm$ 0.03 & 0.119 & 6.1 & 0\farcs65×0\farcs62 (-62.0°) \\
\hline
HD 163296 & HCO$^{+}$ $J = 4 - 3$ & 356.734 & 20.17 $\pm$ 2.02 & 0.119 & 13 & 0\farcs42×0\farcs33 (-52.0°) \\
 & H$^{13}$CO$^{+}$ $J = 3 - 2$ & 260.255 & 0.87 $\pm$ 0.09 & 0.119 & 5.4 & 0\farcs59×0\farcs49 (86.0°) \\
 & N$_2$H$^{+}$ $J = 3 - 2$ & 279.512 & 0.46 $\pm$ 0.05 & 0.119 & 5.1 & 0\farcs55×0\farcs39 (-76.0°) \\
 & N$_2$H$^{+}$ $J = 4 - 3$ & 372.672 & 1.44 $\pm$ 0.15 & 0.119 & 20 & 0\farcs54×0\farcs46 (-72.0°) \\
 & DCO$^{+}$ $J = 4 - 3$ & 288.144 & 1.30 $\pm$ 0.13 & 0.119 & 7.9 & 0\farcs47×0\farcs35 (-79.0°) \\
 & DCO$^{+}$ $J = 5 - 4$ & 360.170 & 2.55 $\pm$ 0.26 & 0.119 & 5.0 & 0\farcs44×0\farcs34 (-58.0°) \\
\hline
LkCa 15 & HCO$^{+}$ $J = 4 - 3$ & 356.734 & 9.76 $\pm$ 0.98 & 0.119 & 18 & 0\farcs53×0\farcs41 (48.0°) \\
 & H$^{13}$CO$^{+}$ $J = 3 - 2$ & 260.255 & 0.39 $\pm$ 0.04 & 0.119 & 5.1 & 0\farcs68×0\farcs48 (-14.0°) \\
 & N$_2$H$^{+}$ $J = 3 - 2$ & 279.512 & 1.60 $\pm$ 0.17 & 0.139 & 3.6 & 0\farcs33×0\farcs24 (-20.0°) \\
 & N$_2$H$^{+}$ $J = 4 - 3$ & 372.672 & 2.17 $\pm$ 0.25 & 0.119 & 24 & 0\farcs40×0\farcs31 (-44.0°) \\
 & DCO$^{+}$ $J = 4 - 3$ & 288.144 & 0.57 $\pm$ 0.06 & 0.129  & 3.0 & 0\farcs30×0\farcs22 (-15.0°) \\
 & DCO$^{+}$ $J = 5 - 4$ & 360.170 & 0.55 $\pm$ 0.06 & 0.119 & 8.2 & 0\farcs68×0\farcs51 (62.0°) \\
\hline
MWC 480 & HCO$^{+}$ $J = 4 - 3$ & 356.734 & 8.87 $\pm$ 0.89 & 0.119  & 6.7 & 0\farcs34×0\farcs26 (17.0°) \\
 & H$^{13}$CO$^{+}$ $J = 3 - 2$ & 260.255 & 0.28 $\pm$ 0.03 & 0.119 & 5.0 & 0\farcs76×0\farcs47 (-7.0°) \\
 & N$_2$H$^{+}$ $J = 3 - 2$ & 279.512 & 0.17 $\pm$ 0.03 & 1.09 & 2.3 & 1\farcs15×0\farcs53 (-32.0°) \\
 & N$_2$H$^{+}$ $J = 4 - 3$ & 372.672 & 0.65 $\pm$ 0.07 & 0.119 & 12 & 0\farcs44×0\farcs36 (8.0°) \\
 & DCO$^{+}$ $J = 4 - 3$ & 288.144 & 1.04 $\pm$ 0.11 & 1.09 & 2.9 & 1\farcs10×0\farcs91 (-7.0°) \\
 & DCO$^{+}$ $J = 5 - 4$ & 360.170 & 1.02 $\pm$ 0.11 & 0.119 & 2.8 & 0\farcs34×0\farcs26 (18.0°) \\
\hline
V4046 Sgr & HCO$^{+}$ $J = 4 - 3$ & 356.734 & 26.38 $\pm$ 2.77 & 0.119 & 9.2 & 1\farcs33×0\farcs58 (-88.0°) \\
 & H$^{13}$CO$^{+}$ $J = 3 - 2$ & 260.255 & 1.00 $\pm$ 0.10 & 0.119 & 5.6 & 0\farcs60×0\farcs49 (87.0°) \\
 & N$_2$H$^{+}$ $J = 3 - 2$ & 279.512 & 3.67 $\pm$ 0.37 & 0.139 & 6.0 & 0\farcs58×0\farcs48 (-77.0°) \\
 & N$_2$H$^{+}$ $J = 4 - 3$ & 372.672 & 3.68 $\pm$ 0.37 & 0.119  & 27 & 1\farcs18×0\farcs78 (-83.0°) \\
 & DCO$^{+}$ $J = 4 - 3$ & 288.144 & 1.56 $\pm$ 0.16 & 0.129 & 4.6 & 0\farcs56×0\farcs45 (-77.0°) \\
 & DCO$^{+}$ $J = 5 - 4$ & 360.170 & 1.69 $\pm$ 0.18 & 0.119 & 5.0 & 0\farcs80×0\farcs51 (-83.0°) \\
\hline
    \end{tabularx}
    \begin{tablenotes}
      \small
      \item \textbf{Notes.} $^{(1)}$ Line fluxes were measured within Keplerian masks. Errors were estimated by applying the same masks to line-free channels and include the nominal 10$\%$ flux calibration uncertainty from ALMA added in quadrature.
    \end{tablenotes}
    \label{tab:fluxes}
\end{threeparttable}
\end{table*}

The \hcop\ data from program 2021.1.00138.S are incomplete along the velocity axis due to a spectral setup error, leaving a missing redshifted component in five disks (AS 209, GM Aur, HD 163296, LkCa 15, and MWC 480). For these sources, we assume symmetric emission and estimate the integrated flux by doubling the flux measured from the unaffected half of the data cube. This assumption is validated using the two disks with complete \hcop\ $J=4–3$ coverage, DM Tau and V4046 Sgr, for which the half-spectrum flux differs from the full-spectrum value by only 3–5$\%$. The assumption of symmetric \hcop\ emission is further supported by previous observations of \hcop\ $J =3-2$ in DM Tau, MWC 480, LkCa 15, and GM Aur \citep{guilloteau_2016}, and \hcop\ $J=4-3$ in HD 163296 \citep{mathews13}, none of which exhibit significant asymmetries. When constructing azimuthally averaged radial profiles with \texttt{GoFish}, we restrict the polar angle to the region with complete \hcop\ coverage to avoid artificially suppressing emission in affected azimuthal sectors.

\section{Results}\label{sec:results}

Lines with integrated flux values above a 3$\sigma$ level and emission above 3$\times$RMS in at least 5 individual channels are considered detected. Based on these criteria, we detect all of the lines in all of our sources. 

\subsection{Emission Morphologies}

The zeroth-moment maps reveal diverse molecular ion morphologies across the sample (Figure~\ref{fig:mom0}). Several disks exhibit ringed emission, most notably AS~209 in \nthp\ and \dcop. In general, molecular ion emission consists of broad single rings rather than the multiple narrow rings and gaps seen in high-resolution continuum data. Because many continuum substructures are unresolved at the spatial resolution of the line observations, additional unresolved line substructure may also be present. HD~163296 is the main exception, exhibiting multiple rings in several tracers. The clearest example occurs in \dcop, where bright rings at $\sim$49 and 110 au are separated by a gap near 80 au.

 \begin{figure*}
    \centering
    \includegraphics[width=0.94\textwidth]{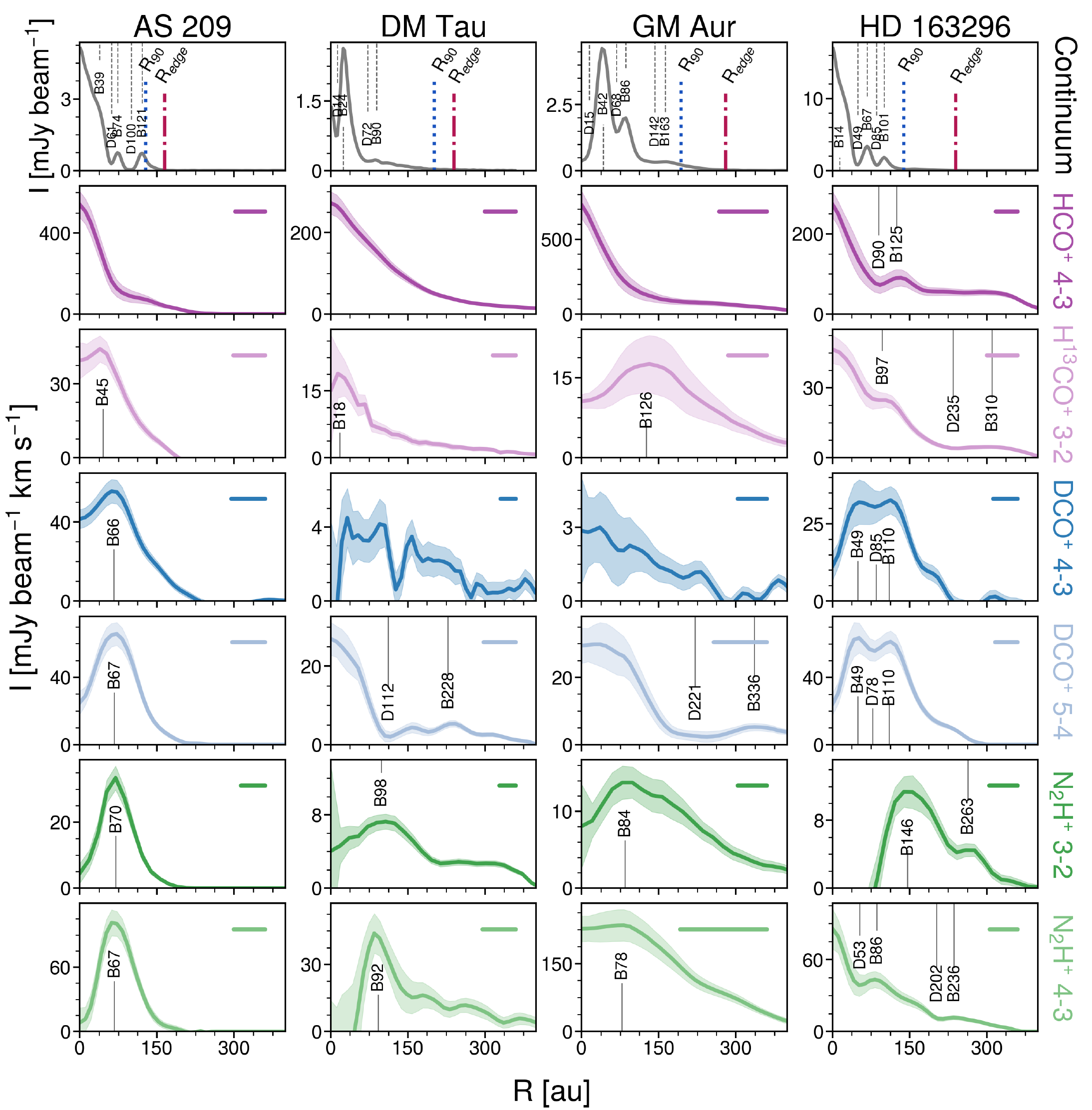}
        \caption{\normalsize{Deprojected radial intensity profiles for all of the lines in our survey, as listed in Table \ref{tab:obs}. A horizontal bar in the upper right corner of each panel represents the FWHM of the synthesized beam. In the top row we include the deprojected radial intensity profiles for high resolution continuum observations from the MAPS and exoALMA Large Programs, at frequencies of 260 GHz and 331 GHz, respectively \citep{law_2021_rad,curone_2025}. We plot the published R$_{90}$ values as well as millimeter dust edge locations (R$_{edge}$), identified by visual inspection, as blue dotted and red dashed-dotted lines, respectively. Several of the MAPS disks have additional low-contrast dust rings and gaps located at $<$ 40 au that were identified by \citet{Huang_2018} and \citet{Huang_2020} but are not plotted here as they were not resolved in the 260 GHz MAPS data. For full details on the characterization of substructures in the continuum profiles see \citet{law_2021_rad} and \citet{curone_2025}. }}
  \label{fig:radprofs}
\end{figure*}

Differences in angular resolution complicate direct comparisons between transitions. For example, the narrow \dcop\ $J=5-4$ ring in MWC~480 is spatially resolved, whereas the lower-resolution $J=4-3$ data appear centrally peaked. Apparent morphological differences between transitions should therefore be interpreted with caution, with greater weight given to the higher-resolution observations. Additionally, different transitions of the same molecular species could trace different emission layers with distinct physical conditions resulting in morphological differences. These effects must be taken into consideration when comparing morphologies of different transitions. 

\subsection{Radial Distributions of Molecular Ions}

\addtocounter{figure}{-1}
 \begin{figure*}
    \centering
    \includegraphics[width=0.75\textwidth]{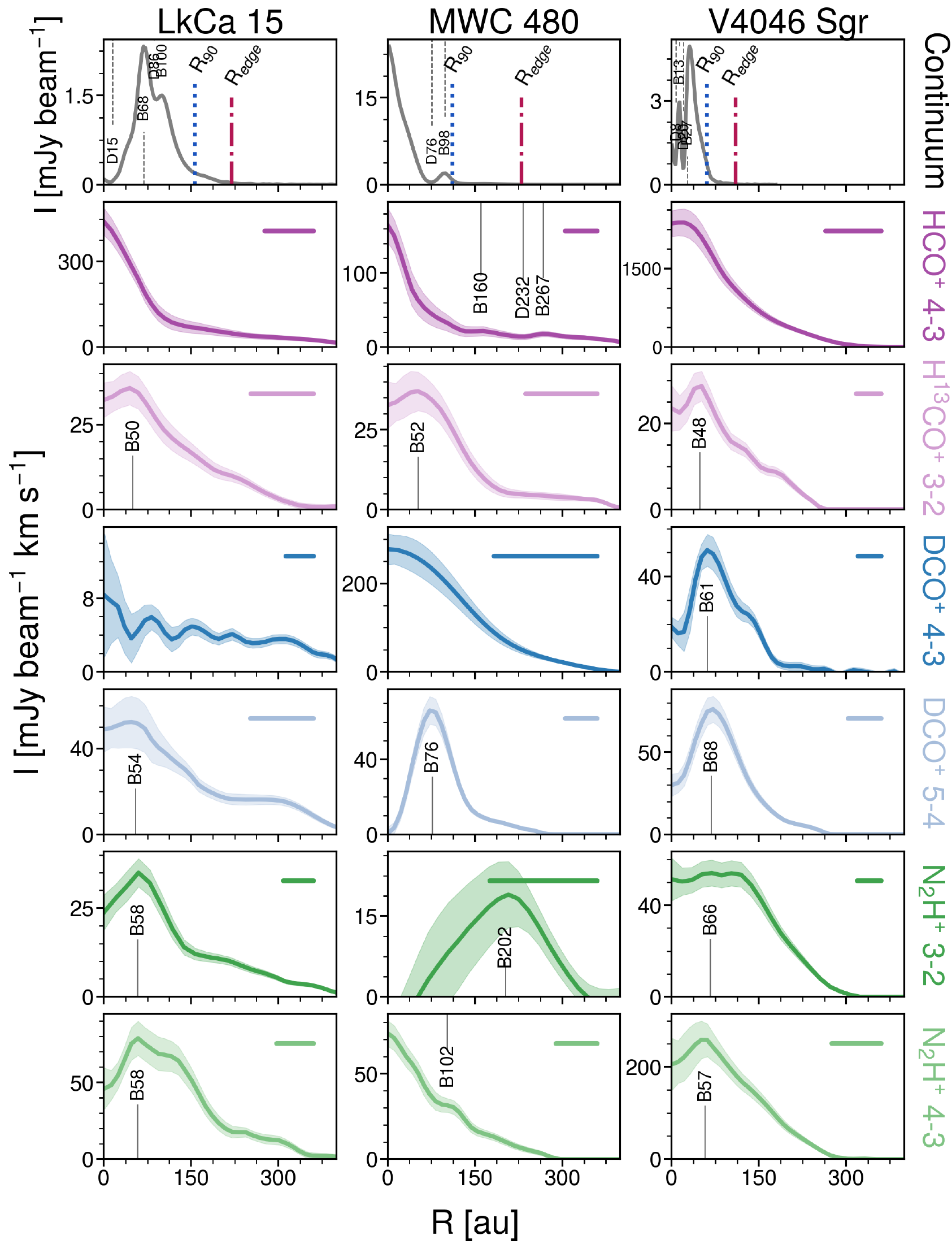}
        \caption{\normalsize{continued.}}
  \label{fig:radprofs}
\end{figure*}

We use the azimuthally-averaged radial intensity profiles to identify substructures (e.g., gaps and rings) in molecular ion emission for all six lines in our survey. We adopt a naming convention similar to the one used to characterize substructures in continuum emission \citep{Huang_2018}, where ``dark" gap features are denoted by the prefix D and ``bright" ring features by the
prefix B. We do not report estimated widths or depths of the features given the varying resolution of data in our sample. Figure \ref{fig:radprofs} shows the full gallery of radial profiles with substructure labels. A complete list of substructure features can be found in Table \ref{tab:ion_substr} in Appendix \ref{app:substr}. Some profiles exhibit deviations that cannot be characterized as rings or gaps, but rather might be considered to be emission ``plateaus” \citep{Huang_2018,Huang_2020}. We only label such features if they clearly correspond to emission features visible in the zeroth moment maps, otherwise we do not label them or include them in our quantitative analysis. 

\hcop\ is bright and centrally peaked in all disks. In most disks the \hcop\ emission drops off smoothly, although DM Tau, GM Aur, LkCa 15, and AS 209 exhibit broad outer emission plateaus. Distinct rings and gaps are observed only in the Herbig disks MWC~480 and HD~163296. Comparison with MAPS data for sources AS 209, HD 163296, GM Aur, and MWC 480 indicates that the \hcop\ $J=4-3$ profiles more closely resemble CO emission than \hcop\ $J=1-0$, consistent with optically thick emission tracing gas structure rather than \hcop\ abundance. In HD~163296, MAPS observations of CO $J = 2 - 1$, $^{13}$CO $J = 2 - 1$, and C$^{18}$O $J = 2 - 1$ exhibit gaps and rings at similar radial locations to those seen in \hcop\ $J = 4 - 3$ \citep{law_2021_rad}. The \hcop\ substructures in MWC 480 could similarly coincide with CO enhancements at $\sim$200 and  $\sim$300 au, however the lower spatial resolution of our observations limits direct comparison.

In contrast to \hcop, \htcop\ generally exhibits central depressions. These are most pronounced in AS 209, GM Aur, and V4046 Sgr, where central depressions are clearly apparent in the moment maps. Rings of \htcop\ are detected in all disks. Six of the seven sources display a single ring outside of the central depression, while HD~163296 exhibits two rings separated by a broad, low-contrast gap. Overall, the \htcop\ $J=3-2$ profiles resemble previously observed ring-like \hcop\ $J=1-0$ emission, although the central depressions are typically shallower. GM~Aur is the notable exception, with centrally peaked \hcop\ but ringed \htcop\ emission.

\dcop\ shows a variety of radial emission morphologies across the sample. Most disks show a single emission ring, although several display additional inner or outer substructure. HD~163296 is the only source with multiple distinct rings, while V4046~Sgr and HD~163296 exhibit extended outer plateaus. DM~Tau and GM~Aur show contrasting morphologies between transitions, with diffuse $J=4-3$ emission and brighter, centrally peaked $J=5-4$ emission. For AS~209, LkCa~15, HD~163296, MWC~480, and V4046~Sgr, the $J=4-3$ and $J=5-4$ profiles are broadly consistent with previously reported $J=3-2$ observations \citet{huang17}.

\nthp\ generally exhibits ringed emission. Rings are identified in all disks. Four disks (AS 209, GM Aur, LkCa 15, and V4046 Sgr) exhibit central depressions followed by isolated bright \nthp\ rings. DM~Tau instead shows broad emission plateaus, particularly in the $J=3-2$ transition. Another notable feature of the \nthp\ profiles is the contrast between the two transitions in the Herbig sources: the \nthp\ $J=3-2$ profiles exhibit central depressions, whereas the $J=4-3$ profiles are centrally peaked. We discuss possible interpretations of these contrasting morphologies in Section \ref{sec:herbigs}.

\subsection{Ion Flux Correlations}

To assess global relationships between ion emission and source properties we look at the disk-integrated ion fluxes. We plot molecular ion fluxes against source properties from Table \ref{tab:source_props} including stellar mass, stellar X-ray luminosity, and mass accretion rate, as well flux quantities including the 0.9 mm continuum flux, $^{13}$CO flux, \hcop\ flux, and \htcop\ flux. We use the 0.9 mm continuum fluxes from this work (Table \ref{tab:cont}) and published $^{13}$CO $J = 2 - 1$ fluxes from \citet{oberg_maps_2021}, \citet{long_22}, and \citet{leemker_2022}. Fluxes are scaled to a common distance of 160 pc. 

\begin{figure*}
    \centering
    \includegraphics[width=0.75\textwidth]{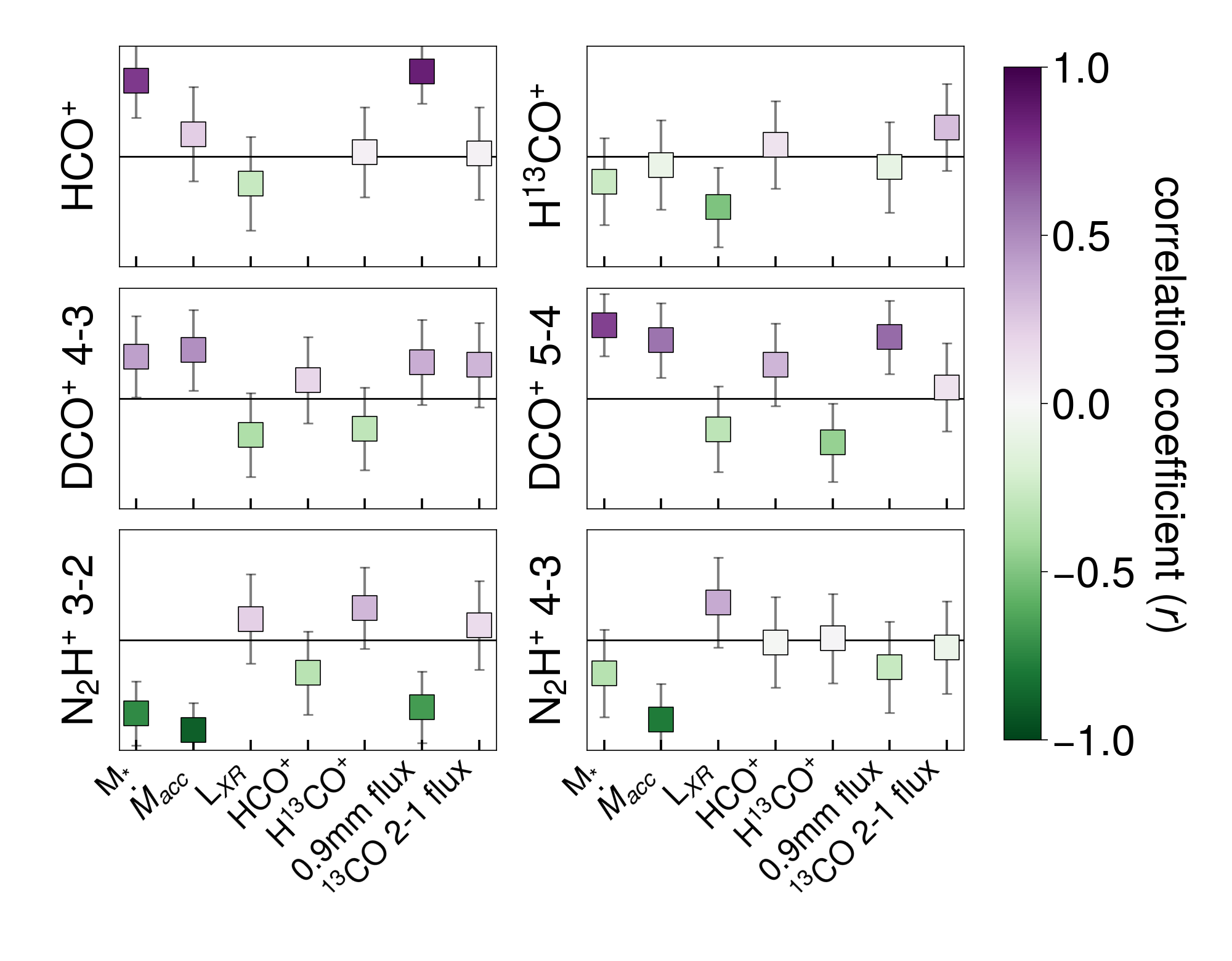}
    \caption{\normalsize{Panels show the median correlation coefficients for each molecular line (labeled to the left of each panel) and different source parameters, with error bars representing one standard deviation. Marker colors map to the colorbar representing the correlation coefficient, with positive values in purple and negative values in green. The horizontal axis is labeled by the source parameters tested for correlations, including stellar mass, mass accretion rate, stellar X-ray luminosity, \hcop\ flux, \htcop\ flux, 0.9 mm continuum flux, and $^{13}$CO flux from left to right.}}
  \label{fig:linmix}
  \end{figure*}

We evaluate the robustness of correlations using the Bayesian linear regression fitting tool \texttt{linmix}\footnote{\url{https://github.com/jmeyers314/linmix}}, which follows the methods presented in \citet{kelly_2007}. Median correlation coefficients ($r$) for all fits are presented in Figure \ref{fig:linmix}, while a figure showing the fluxes, median fits, and posterior distributions can be found in Appendix \ref{app:fluxcorr} Figure \ref{fig:fluxcorr}. Six strong correlations are found between line fluxes and source parameters:

\begin{itemize}
    \item High positive correlations ($r \ >$ 0.7)
    \begin{itemize}
        \item[1.] \hcop\ $4-3$ and stellar mass
        \item[2.] \hcop\ $4-3$ and continuum flux
        \item[3.] \dcop\ $5-4$ and stellar mass
    \end{itemize}
    \item High negative correlations ($r \ <$ -0.7)
    \begin{itemize}
        \item[1.] \nthp\ $3-2$ and stellar mass
        \item[2.] \nthp\ $4-3$ and mass accretion rate
    \end{itemize}
    \item Very high negative correlations ($r \ <$ -0.9)
    \begin{itemize}
        \item[1.] \nthp\ $3-2$ and mass accretion rate
    \end{itemize}
\end{itemize}

\hcop\ $4-3$ flux correlates positively with both stellar mass and continuum flux. Stellar mass is also positively correlated with \dcop\ $5-4$ and negatively correlated with \nthp\ $3-2$, however we note that the correlations/anti-correlations with stellar mass are largely influenced by the two higher-mass Herbig sources and relationships among the T-Tauri sources are otherwise relatively flat (Figure \ref{fig:fluxcorr}). Both \nthp\ lines are negatively correlated with stellar mass accretion rate. The most robust correlation we find is that between the $3-2$ transition of \nthp\ and stellar mass accretion rate, which has a correlation coefficient of -0.89 with a standard deviation of 0.27. The implications of this relationship are discussed further in Section \ref{sec:macc}.

\section{Discussion}\label{sec:disc}

We now present interpretations and implications of the observed distribution of ions in our sample. We first discuss relationships between molecular ions and dust disk structure. We then discuss interpretations of patterns in ion morphologies across the sample, including T-Tauri sources with radially coincident ion rings and Herbig Ae sources. We then discuss factors that may influence the observed ion emission in this sample, including X-ray and cosmic ray ionization, magnetic fields, winds, and accretion.

\subsection{Relationship Between Ion Substructures and Continuum Features}\label{disc:cont}

\begin{figure*}
    \centering
    \includegraphics[width=0.99\textwidth]{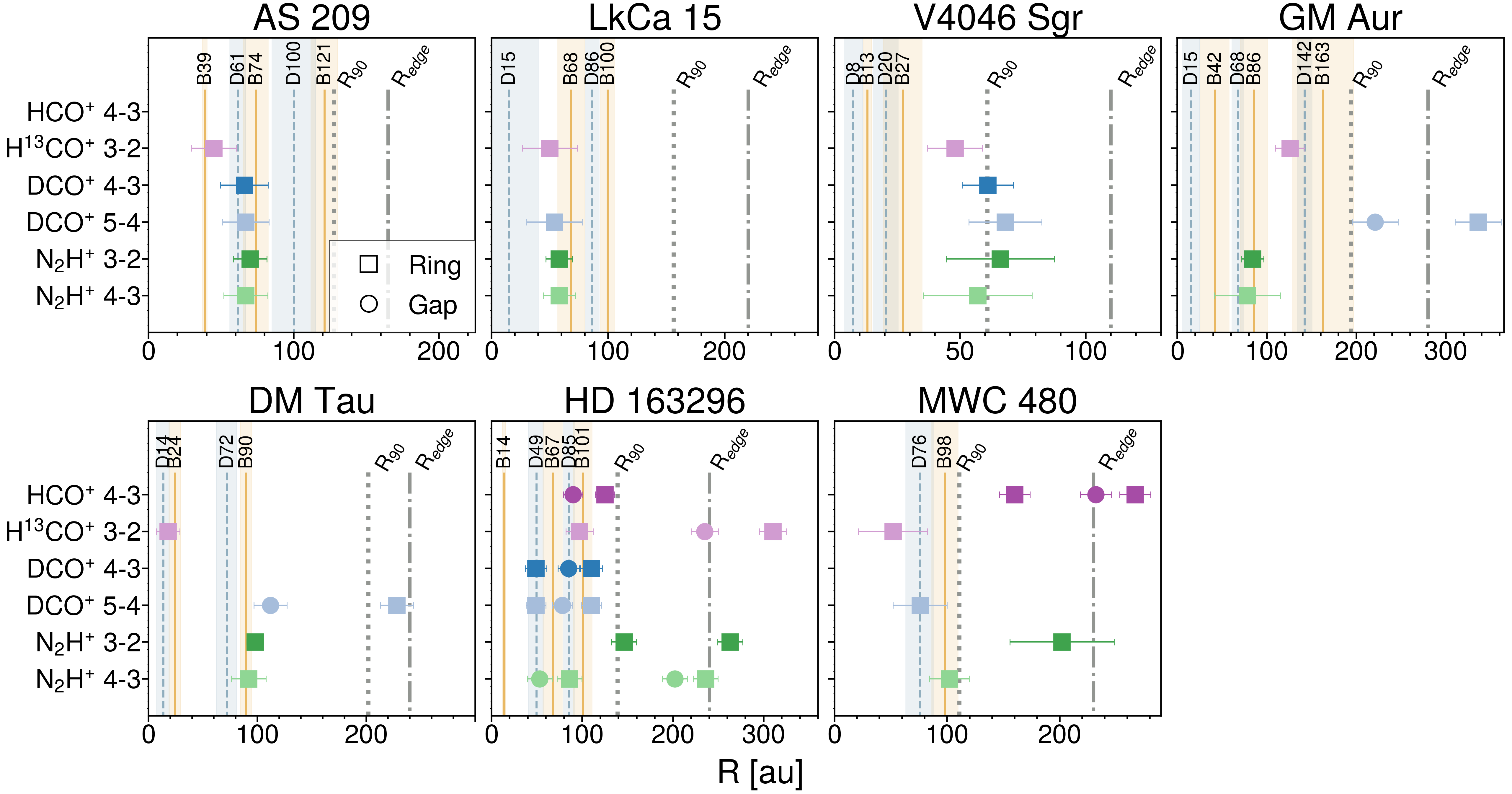}
    \caption{\normalsize{Ion substructure locations from this work are plotted as square and circular markers for rings and gaps, respectively, with colors corresponding to the radial profiles in Figure \ref{fig:radprofs}. Continuum substructure, including rings (solid orange) and gaps (dashed blue), are plotted as vertical lines in the background with shading representing the estimated width of the dust substructures. R$_{90}$ and the visual millimeter dust edge R$_{edge}$ are plotted as grey dotted and dashed-dotted lines, respectively.} }
  \label{fig:contcomp}
  \end{figure*}

We compare the morphologies of ion emission to continuum data from the MAPS and exoALMA Large Programs \citep{law_2021_rad,Sierra21,curone_2025}. We select the highest resolution continuum data from each program, corresponding to frequencies of 260 GHz for MAPS and 331 GHz for exoALMA. The selected continuum maps and radial profiles are shown in the top rows of Figures \ref{fig:cont} and \ref{fig:radprofs}, respectively. There are multiple ways of defining and measuring dust disk sizes \citep{tripathi_2017, Long_2018}. We consider two definitions of continuum disk size in our analysis: R$_{\rm{90}}$, which is defined as the radius containing 90$\%$ of the continuum flux, and R$_{\rm{edge}}$, defined as the outermost edge of the continuum disk. We use published R$_{\rm{90}}$ values from \citet{law_maps_err} and \citet{curone_2025}. We measure R$_{\rm{edge}}$ via visual inspection of the continuum radial profiles. In Figure \ref{fig:contcomp}, we plot ion substructure locations over continuum substructure, R$_{\rm{90}}$, and R$_{\rm{edge}}$ locations.

All of the sources have some ion substructures coincident with continuum substructures, but relationships are diverse across sources and ion species. Three of the disks--AS 209, LkCa 15, and V4046 Sgr--have ion substructures in multiple species clustered within a relatively narrow range of radii. Most of these clusters fall between a dust gap and dust ring or vice versa. The ion rings in AS 209 are all located within the dust gap between $\sim$40-70 au. All of the ion rings in LkCa 15 are located interior to the dust ring at 68 au. In V4046 Sgr, the ion rings are clustered near R$_{\rm{90}}$ at $\sim$60 au. 

In DM Tau, the \htcop\ ring lies interior to the first dust ring at 24 au, while both \nthp\ rings coincide with a low-contrast dust ring at 90 au. GM Aur exhibits \nthp\ rings overlapping with a dust ring at 86 au. In HD 163296, most ion rings in the inner 100 au overlap somewhat with the dust gaps and rings in that region. However, because the dust substructures of interest are separated by less than 20 au, higher resolution ion observations are required to clearly determine any relationships between ion and continuum features in this disk. In MWC 480, \dcop\ $J = 5-4$ and \nthp\ $J = 4-3$ rings are coincident with the dust gap and ring at 76 and 98 au respectively.

\subsubsection{Millimeter Dust Edge}

Previous observations point to a tentative correlation between molecular rings and the millimeter dust edge \citep{oberg15,guzman2015,bergin16}. It has been shown through modeling that the removal of dust mass could make CO in the outer disk more susceptible to thermal or non-thermal desorption by allowing the disk to become warmer or simply more permeable to external UV radiation \citep{oberg15,cleeves16b}. This mechanism could explain the presence of rings in CO-bearing molecules, such as \dcop\ and \hcop, near dust edges. 

No clear pattern emerges for the five sources with some ion rings coincident with either R$_{\rm{90}}$ or the visual millimeter dust edge. In HD 163296, outer disk \nthp\ rings overlap with R$_{\rm{90}}$ and R$_{\rm{edge}}$. The \htcop\ gap, corresponding to the \htcop\ emission break around 200 au first identified by \citet{huang17}, also coincides with R$_{\rm{edge}}$. In MWC 480, there is a gap and an outer ring of \hcop\ at the edge of the dust disk, perhaps evidence of increased gas-phase CO exterior to the dust edge. In V4046 Sgr, the \dcop\ and \nthp\ rings are all coincident with the dust edge at 60 au. Both DM Tau and GM Aur have diffuse rings of \dcop\ in their outer disks near the dust edge. 

Our small sample does not reveal a clear trend--different molecules are present at the edge of the dust disk, as traced by either R$_{\rm{90}}$ or R$_{\rm{edge}}$. However, the fact that there are rings and emission plateaus seen in both CO-bearing (\hcop, \dcop) species and non-CO-bearing species (\nthp) coincident with the dust edge in these sources suggests that these features could be tracing enhanced ionization beyond the dust edge, perhaps due to increased UV radiation \citep{walsh_13, gross_25} or permeability to external CRs, rather than CO desorption alone. 

\subsubsection{Transition Disk Cavities}

Four of the disks in our sample are known transition disks \citep[cavity size $>$ 15 au]{vandermarel_2023}. LkCa 15 is the only transition disk with multiple ion rings inside of the dust cavity. LkCa 15 is also the transition disk with the largest cavity at $\sim$70 au. It is thus possible that the presence of rings in all three ion species within the cavity is related to increased ionization in the cavity. DM Tau has an \htcop\ ring inside the cavity but the other ions peak exterior to the dust cavity. Forward modeling of ion profiles in DM Tau suggested that the disk may be subject to a high rate of ionization inside the 20 au cavity, possibly due to ionization by stellar energetic particles \citep{long24, brunn_24}. This scenario could explain the presence of the \htcop\ ring within the cavity of DM Tau, its location set by the increased ionization degree in the cavity, and cold ion rings further out in the disk, if their locations are instead set by chemical effects. In GM Aur and V4046 Sgr, all of the ion rings are located exterior to the cavity, and the ion substructure locations do not appear to be related to the cavity. Overall, the four transition disks in our sample display a variety of ion emission morphologies and hint at cavity-related ionization effects in LkCa 15 and DM Tau.

Two transition disks, DM Tau and GM Aur, exhibit centrally-peaked \dcop\ $J =5-4$ profiles. If the \dcop\ in these sources is centrally peaked, it may indicate distinct chemical effects. One possible explanation is that higher dust cavity temperatures and/or X-ray permeability in the cavities of transition disks could lead to production of \dcop\ predominantly via C\htdp\ instead of \htdp, as C\htdp\ is more abundant than \htdp\ at higher temperatures \citep{favre_2015}. However, it is unclear why this effect would be evident in only two of the four transition disks in the sample. If there are unresolved \dcop\ rings in DM Tau or GM Aur, they would be located at $<$ 60 au and 100 au, respectively, interior to \nthp\ rings in DM Tau and \nthp\ and \htcop\ rings in GM Aur. In that case, DM Tau might better follow the expected chemical trend of nested rings of \htcop, \dcop, and \nthp\ moving outward as they probe chemistry at colder temperatures from the warm surface to midplane. Higher resolution observations of these sources will help to elucidate the distribution of \dcop\ and clarify how similar or dissimilar their ion morphologies are to each other and other T-Tauri systems.

\subsection{Interpreting Ion Morphologies Across the Sample}

In order to understand the ionization environments of protoplanetary disks from observations of molecular ions, we must be able to reliably interpret the observed molecular line fluxes and emission morphologies of these species. This interpretation is complicated by a variety of factors and processes that influence the intensity and distribution of ion emission. In addition to the actual degree of ionization in the disk, ion emission may be impacted by non-LTE and excitation effects, chemical variations such as CO depletion changing the abundance of molecular ions, changes in temperature or density, and more \citep{pavlyuchenkov_2007,anderson_2019}. 
Taking these effects into account has typically required detailed modeling of individual sources, and it can therefore be challenging to generalize results. In this section, we focus our analysis on observational trends in the radial intensity profiles across our sample and how they compare to previous modeling efforts in the literature.  

\subsubsection{Radial Profiles}

\begin{figure*}
    \centering
    \includegraphics[width=0.95\textwidth]{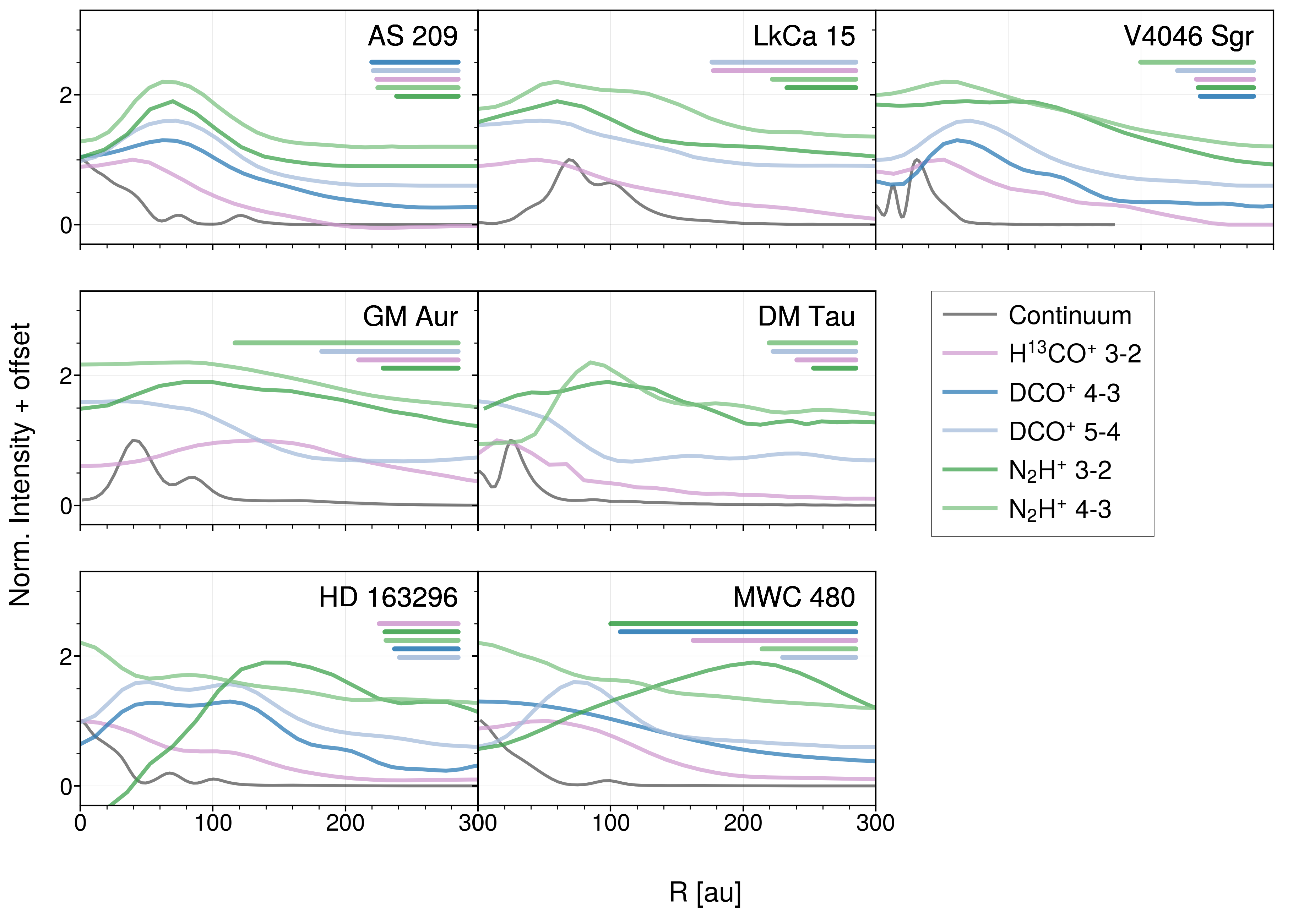}
    \caption{\normalsize{Radial intensity profiles for molecular ions and continuum emission normalized to a maximum value of 1, with vertical offsets added in increments of 0.3 to make individual profiles more visible. Beam sizes for the molecular ion profiles are plotted as horizontal bars in the same color as the corresponding molecular tracer. \textit{Top row:} T-Tauri sources with radially coincident emission rings. \textit{Middle row:} T-Tauri sources with varied ion emission morphologies. \textit{Bottom row:} Herbig sources with varied ion emission morphologies.}}
  \label{fig:pattern}
  \end{figure*}

Figure \ref{fig:pattern} shows the normalized radial ion and continuum emission profiles for our sample. The radial emission profiles reveal a distinction between T-Tauri sources with radially coincident emission rings in multiple optically thin ion species vs. those with varied morphologies across different ion species. The two Herbig sources also have varied emission morphologies across ion species and exhibit interesting similarities across the two sources, namely centrally peaked \nthp\ $J=4-3$ emission that isn't observed in the T-Tauri sources. 

A subset of three T-Tauri sources (AS 209, LkCa 15, and V4046 Sgr) exhibit overlapping, radially coincident emission rings across multiple molecular ion species. In these disks, the optically-thin molecular ions have ring-like distributions peaking within a small range of radii. The maximum radial separation between ion ring peaks for LkCa 15, V4046 Sgr, and AS 209 are 8 au, 20 au, and 25 au, respectively. That the ion rings are so tightly spaced in these sources may indicate that the ion emission morphology is dominated by a global condition, such as gas density or an ionization gradient.

The second subset of T-Tauri disks, DM Tau and GM Aur, do not exhibit the same tight radial correlations in molecular ion distribution across multiple species. Rather, different ions exhibit different distributions. Ion rings, when present, peak across a wider range of radii than seen in the other subset of T-Tauri disks. The lack of correlation among molecular ion distributions may indicate that their chemistry may be dominated by smoother gradients in the local conditions. This is aligned with what we expect theoretically--that each molecular ion species, sensitive to the chemistry and underlying physical conditions, traces a different vertical zone of the disk. DM Tau and GM Aur may therefore be preferred candidates for constraining 2D ionization structures. 

The T-Tauri sources with strongly overlapping ion rings do still exhibit small offsets in ring peaks that follow the expected chemistry-driven structure, with the largest shift being \htcop\ peaking radially closer to the stars than \dcop\ or \nthp, possibly tracing the warmer regions where CO is still abundant in the gas phase. The other rings' peaks are shifted slightly outward-deuterated \dcop, perhaps tracing lower temperatures, and finally \nthp, which is expected to trace the cold reservoir beyond the CO snowline. Nonetheless, these shifts are substantially smaller than those in GM Aur or DM Tau (the uncorrelated sources), the latter having rings of \htcop\ and \nthp\ separated by more than 70 au. 

While our observations map the radial distributions of molecular ions, we do not directly measure their vertical distribution and instead rely on theoretical expectations to interpret the radial profiles in a 2D context. It has recently been demonstrated that observations with sufficient signal-to-noise and resolution can be used to measure the emitting layers of molecular tracers, but so far, molecular ions have only been mapped vertically in a few systems \citep[][]{Huang_2020, paneque_carreno_2023, Law24, Foucher25}. However, direct constraints on the vertical distribution of molecular ion emission are needed in addition to detailed modeling to more accurately interpret observed differences in ion emission morphologies across sources. 

\subsubsection{Two Emerging Categories of T-Tauri Cosmic-ray Ionization Environments}

Coherent emission rings in \nthp\ and \htcop\ are surprising, as both chemical models \citep[e.g.,][]{Aikawa_2021} and synthetic observations \citep[e.g.,][]{cleeves14a} find that under standard chemical assumptions for disks exposed to a uniform incident cosmic ray ionization rate (CRIR), different ions' abundance distributions or emission intensities peak at different radial and vertical locations. Specifically, \citet{cleeves14a} examined models with varying global cosmic ray ionization rates, finding the resulting emission profiles have distinct, widely separated ($>$100 au) rings for the molecules discussed here. 

More recently, observations of radially coincident \nthp\ and \htcop\ emission rings have been interpreted in the context of a global ionization gradient \citep{Seifert_2021}. In this work, the authors find the IM Lup disk is best modeled with a sharp increase in the CRIR at a ``swap radius" of 100 au. This was required to reproduce observations of \htcop\ and \nthp\ where both species have central deficits and molecular rings peaking at $\sim$160 au \citep{Seifert_2021}. 

That work speculated the steep increase in the CRIR at 100 au could be the result of a T-Tauriosphere--a stellar wind analogous to the Sun’s heliosphere--blocking galactic CRs from reaching the inner disk \citep{cleeves13a}. The presence of a T-Tauriosphere was first proposed to explain the low CRIR across the full extent of the TW Hya protoplanetary disk \citep{cleeves15_twhya}. \citet{Seifert_2021} suggest that the presence of a cosmic ray gradient in IM Lup but not TW Hya might be related to the system's evolutionary stage. They suggest that a young system like IM Lup ($\sim$0.5 Myr) might have its wind constrained by the presence of a remnant envelope or otherwise unable to encompass IM Lup's 600~au gas disk. 

If cosmic ray exclusion is related to evolutionary state, this is not apparent from our data since AS 209, LkCa 15, and V4046 Sgr range in age from 1.6 Myr up to $>$ 5 Myr. Instead, if the wind scenario holds, it is more likely that a combination of factors will set its efficiency, such as amount of surrounding ISM and the momentum carried by the wind itself. 

Another data point comes from an in-depth study of DM Tau, which found this disk's chemistry is best fit by a reduced CRIR in the outer disk and a high ionization rate in the inner disk \citep{long24}. \citet{long24} suggest that DM Tau may undergo CR modulation, possibly by a stellar wind, across the full disk extent, while the inner disk is subject to a boost in ionization by stellar energetic particles. The more varied ion morphologies we observe in both DM Tau and GM Aur are compatible with the presence of a T-Tauriosphere across the full disk extent. In this scenario we would expect the ion morphologies to be more influenced by local variations due to chemistry or underlying physical structure rather than a sudden change in ionization. 

In the context of published models of the IM Lup, TW Hya, and DM Tau ionization environments, our observations of four additional T-Tauri sources hint at a potential dichotomy between two types of T-Tauri ionization environments. The first are systems where a sharp radial increase in ionization, possibly at the edge of a T-Tauriosphere, results in radially coincident ion emission rings. The second category of systems, those that exhibit more varied ion morphologies, could represent systems where the CRIR is constant or smoothly varying across the disk, possibly because the system lacks a significant wind component or because the wind encompasses the entire disk. While in-depth characterization of these systems requires detailed physical and chemical modeling, resolved ion observations point toward differences among T-Tauri ionization environments. 

\subsubsection{Centrally-Peaked \nthp\ in Herbig Sources}\label{sec:herbigs}

The Herbig sources in our sample exhibit a discrepancy in emission morphology between the two \nthp\ lines that is not seen in the T-Tauri sources. In both Herbig sources, the \nthp\ $J = 3-2$ profiles have central cavities with peaks further out in the disk whereas the \nthp\ $J = 4-3$ profiles are centrally peaked with secondary ring-like features, possibly corresponding to the peaks in the $J = 3-2$ profiles. The T-Tauri sources in our sample exhibit distinctly ringed emission following similar distributions in both \nthp\ $3-2$ and $4-3$, suggesting that both lines trace similar emission reservoirs. Excitation conditions derived for \nthp \ $4-3$ in the GG Tau triple stellar system showed that the line traces a layer close to the midplane with an average temperature of 12 K \citep{parashmoni_2024}. While observations of these two transitions of \nthp\ in T-Tauris appear to trace similar emission reservoirs close to the midplane, this may not be the case for the Herbigs observed here.

The distinct morphologies observed in the two Herbig sources suggest that the different \nthp\ transitions, with respective upper energy levels of 26.8 K and 44.7 K, could probe different \nthp\ emitting regions in these disks. One interpretation is that the \nthp\ $J = 4-3$ transition traces a warm \nthp\ ``surface layer". This surface layer could be the result of either enhanced ionization or CO destruction via photodissociation, which would allow \nthp\ to form efficiently in higher temperature surface regions or closer to the central star \citep{vanthoff17}. 

Previous studies constraining the location of the CO snowline in HD 163296 found good agreement between the location of the inner edge of the \nthp\ ring and the inferred location of the CO snowline using observations and column density modeling of C$^{18}$O and \nthp\ $J=3-2$ \citep{qi_2015}. In \citet{qi_2015}, this constraint came from visibility-domain modeling of the $J=3-2$ data, and the \nthp\ emission peak can lie well exterior to the snowline even when the inner edge is set by CO freeze-out. These results may indicate that there is no \nthp\ surface layer in HD 163296 since there is no offset between the inferred CO column density drop and the \nthp\ emission peak. However, the \nthp\ $J = 4-3$ profiles challenge this interpretation as they suggest that there may be abundant \nthp\ in the inner disks of both HD 163296 and MWC 480. There may be other explanations for the inner disk \nthp\ emission besides the presence of a surface layer formed by photodissociation of CO. For example, it is possible that the differing \nthp\ $J = 3-2$ and $J = 4-3$ morphologies are driven by excitation and radiative transfer effects (including optical depth), with the higher-J line emission more strongly weighted toward warmer, denser gas in the inner disk and potentially suppressed in the outer disk.

\subsection{Other Factors Influencing the Distribution of Ions}

The sources in our sample vary in stellar mass, mass accretion rate, and X-ray luminosity. In this section, we explore how these properties might impact the observed ion emission. 

\subsubsection{X-ray Ionization Environments}\label{sec:xrs}

\nthp\ is sensitive to the rate of ionization from both X-rays and cosmic rays \citep{walsh_12, cleeves15_twhya, Seifert_2021, long24}. The Herbig sources in our sample have the lowest \nthp\ $J = 3-2$ fluxes, possibly due to the fact that Herbigs are weaker X-ray emitters than T-Tauri sources. However, this distinction is not consistent in the $J = 4-3$ transition where Herbigs and T-Tauris span a similar range of fluxes. V4046 Sgr is somewhat of an outlier with relatively high \nthp\ fluxes, and the distribution of \nthp\ flux is fairly flat if it is excluded. The relatively high fluxes in V4046 Sgr may be related to the binary nature of this system, as irradiation from two central stars could yield a larger emitting area than the single-star systems. Detailed modeling of the observed emission profiles will be necessary to fully constrain the rate of ionization from both cosmic rays and X-rays, and we leave detailed modeling of the full sample to future work.

All of the sources in our sample have X-ray luminosity constraints that we compare with molecular ion fluxes in Figure \ref{fig:fluxcorr} (Appendix \ref{app:fluxcorr}), however our data reveal no obvious relationships between ion fluxes and X-ray luminosity (Figure \ref{fig:linmix}). Both \nthp\ and \hcop\ have been shown to be sensitive to X-ray variability, which could impact the observed fluxes and subsequent comparisons between sources \citep{cleeves17, waggoner_2022}. While molecular ions may trace time-dependent chemistry related to X-ray flares, chemical models show that these events do not cause a lasting change in abundance over time, with \hcop\ and \nthp\ returning to steady state in 20 days and 5 days, respectively \citep{waggoner_2022}. Observations taken during a flare could have higher fluxes, leading to over-estimations on the degree of ionization, deviations from real trends, or false trends. Larger samples and multi-epoch observations will help mitigate the impact of variability in ionization studies.

\subsubsection{Stellar Mass and Accretion Rates}\label{sec:macc}

Our sample spans stellar masses ranging from 0.5--2.1 \solarmass\ and accretion rates ranging from $\sim$10$^{-9}$--10$^{-7}$ \solarmass\ yr$^{-1}$. The higher mass Herbig sources in our sample (HD 163296 and MWC 480) exhibit centrally peaked \nthp\ $J = 4-3$ profiles as discussed in Section \ref{sec:herbigs}. In addition to the centrally peaked \nthp $J = 4-3$, the Herbigs are the only sources in our sample to exhibit rings and gaps in \hcop\ $J = 4-3$. These features could be tracing rings and gaps of gas in the outer disk or changes in \hcop\ molecular abundance if the emission enters the optically thin regime in the outer disk. Differences in \hcop\ emission between Herbig and T-Tauri sources have been identified by previous observations. In a chemical survey of five Herbig AeBe disks, \citet{pegues_23} found that \hcop\ flux decreased for Herbig disks relative to T-Tauri disks. They proposed two potential explanations for this trend: 1) \hcop\ is radially coincident with CO in TTs but not in Herbigs, resulting in \hcop\ tracing different emitting layers; 2) Herbigs are weaker X-ray emitters and are thus less ionized than T-Tauris. 

The strong negative correlation between \nthp\ $3-2$  and stellar mass could be interpreted as evidence of weak X-ray ionization in Herbigs. However, we also report positive correlations between stellar mass and two other molecular ions, \hcop\ $4-3$ and \dcop\ $5-4$. It is possible that the \hcop\ and \dcop\ lines are more optically thick than the \nthp\, and thus less sensitive to the underlying ionization conditions, or that \nthp\ is simply more sensitive to X-ray ionization processes. Overall, the ion fluxes in our sample alone do not suggest that Herbigs are significantly less ionized than T-Tauri sources.

The anti-correlation between \nthp\ and mass accretion rate suggests that systems undergoing more mass accretion have lower \nthp\ fluxes. Higher mass accretion rates are associated with increased FUV energies \citep{johnskrull_2000,hinton_2022}, which may have attendant effects on the \nthp\ abundance. \citet{gross_25} found that increasing the external UV radiation field in disk models from 1 to 100 G$_{0}$ resulted in a factor of 5 decrease in \nthp\ column density in the outer disk ($>$ 100 au) due to the increased photodesorption of CO. It is possible that higher UV fluxes associated with higher accretion rates could have a similar effect on the \nthp\ chemistry, enhancing CO photodesorption and destroying \nthp.

\section{Conclusions}\label{sec:conc}

We present the largest survey of resolved observations of molecular ions to date in a sample of seven protoplanetary disks. We observed and analyzed \hcop\ $J=4-3$, \htcop\ $J=3-2$, \dcop\ $J=4-3$ and $J=5-4$, and \nthp\ $J=3-2$ and $J=4-3$ lines toward the protoplanetary disks around AS 209, DM Tau, GM Aur, HD 163296, LkCa 15, MWC 480, and V4046 Sgr. Our main findings are as follows:

\begin{itemize}
\itemsep0em
    \item \hcop, \htcop, two \dcop\ lines, and two \nthp\ lines are detected above 3$\sigma$ in all seven of the disks.
    \item Many of the molecular ions exhibit substructure in their radial emission profiles. The most common ion distributions are central depressions followed by a single ring, but we also see multiply-ringed emission, emission plateaus, and gaps. 
    \item We do not find robust trends among ion and continuum distributions in the full sample. This may indicate that the underlying dust structure does not have a strong influence on molecular ion emission, or rather that it is simply difficult to determine the strength of it's influence on ionization chemistry from observations alone.  
    \item The molecular ion emission morphologies hint at a dichotomy between two types of T-Tauri ionization environments, where the presence of radially coincident ion rings may be evidence of a global change in ionization that strongly influences the ion emission morphology across all ionized species. 
    \item The Herbig sources exhibit centrally-peaked \nthp\ $J = 4-3$ which we speculate traces an \nthp\ surface layer that is either highly ionized or subject to enhanced CO destruction due to their strong UV fields.

\end{itemize}

This work reveals new hints of spatial and chemical correlations among ion tracers, but larger sample sizes are needed in order to make statistically significant conclusions. Our sample is also biased towards large, bright disks. In reality most protoplanetary disks, including our own Solar System's protosolar disk, are more compact and faint than those targeted here \citep{Ansdell_2016,Williams_2019}. Large, unbiased molecular ion surveys are needed in order to truly understand the distribution of ionization and its impact on the chemical and physical evolution of the broader protoplanetary disk population. 

\facilities{ALMA}
\software{CASA \citep{casa_mcmullin_2007,CASA_22}, \texttt{bettermoments} \citep{teague_2018_bm}, \texttt{GoFish} \citep{GoFish}, \texttt{keplerian\textunderscore mask} \citep{teague_2020_kep}.}

\begin{acknowledgments}
We thank the anonymous referee for their thoughtful comments which greatly improved the manuscript. 

This paper makes use of the following ALMA data: ADS/JAO.ALMA $\#$2012.1.00681.S, ADS/JAO.ALMA $\#$2013.1.00226.S, ADS/JAO.ALMA $\#$2015.1.00657.S, ADS/JAO.ALMA $\#$2015.1.00678.S, ADS/JAO.ALMA $\#$2016.1.00956.S, ADS/JAO.ALMA $\#$2018.1.01055.L, ADS/JAO.ALMA $\#$2019.1.00379.S, ADS/JAO.ALMA $\#$2021.1.00138.S, ADS/JAO.ALMA $\#$2021.1.01123.L. ALMA is a partnership of ESO (representing its member states),
NSF (USA) and NINS (Japan), together with NRC (Canada), MOST and ASIAA (Taiwan), and KASI (Republic of Korea), in cooperation with the Republic of Chile. The Joint ALMA Observatory is operated by
ESO, AUI/NRAO and NAOJ. The National Radio Astronomy Observatory is a facility of the National Science Foundation operated under cooperative agreement by Associated Universities, Inc. As ALMA users we are grateful to the Atacameño (Likan Antai) Elders--the first astronomers of Atacama--whose wisdom and knowledge we continue to learn and benefit from. This research has made use of data and/or software provided by the High Energy Astrophysics Science Archive Research Center (HEASARC), which is a service of the Astrophysics Science Division at NASA/GSFC. The views expressed in this document are those of the author(s) and do not reflect the official policy or position of the U.S. Naval Academy, the Department of the Navy, the Department of War, or the U.S. Government. 

DEL and LIC acknowledge support from NASA ATP 80NSSC20K0529. DEL also acknowledges support from the Virginia Space Grant Consortium, and the Jefferson Scholars Foundation. LIC also acknowledges support from NSF grant no. AST-2205698, the David and Lucille Packard Foundation, and the Research Corporation for Scientific Advancement Cottrell Scholar Award. Support for CJL was provided by NASA through the NASA Hubble Fellowship grant No. HST-HF2-51535.001-A awarded by the Space Telescope Science Institute, which is operated by the Association of Universities for Research in Astronomy, Inc., for NASA, under contract NAS5-26555.

\end{acknowledgments}

\appendix 
\section{Observational details and Band 7 continuum observations.}
\label{sec:app_obsdat}
\FloatBarrier
Table \ref{tab:alma_details_new} lists the ALMA observation details for the datasets presented here for the first time. Table \ref{tab:cont} lists details for the Band 7 continuum observations.

\begin{deluxetable}{lccccccc} 
	\tabletypesize{\footnotesize}
	\tablecaption{2019 \& 2021 ALMA observations \label{tab:alma_details_new}}
	\tablecolumns{8} 
	\tablewidth{\textwidth} 
	\tablehead{
		\colhead{Source}           &
            \colhead{Date}       &
		\colhead{\# Ant.}        &
		\colhead{Baselines}              & 
		\colhead{Total on-source }              &
		\colhead{Bandpass cal.}                        & 
		\colhead{Phase cal.}                        & 
		\colhead{Flux cal.}                        \\
		\colhead{} & 
		\colhead{} & 
            \colhead{}  & 
		\colhead{ (m)} & 
		\colhead{ time (min)} &
		\colhead{} & 
		\colhead{} &
		\colhead{}
		}
\startdata
\multicolumn{8}{c}{2019.1.00379.S} \\
\hline
DM Tau & 10/17/2019 & 43 & 15 -- 740 & 37 & J0423-0120 & J0510+1800 & J0423-0120 \\
GM Aur & 03/27/2021 & 41 & 14 -- 1240 & 35 & J0750+1231 & J0438+3004 & J0750+1231 \\
 & 07/31/2022 & 41 & 15 -- 1301 & 35 & J0510+1800 & J0438+3004 & J0510+1800 \\
\hline
\multicolumn{8}{c}{2021.1.00138.S} \\
\hline
AS 209 & 01/04/2022 & 42 & 14 -- 976 & 25 & J1517-2422 & J1657-2004 & J1517-2422 \\
GM Aur & 08/19/2022 & 44 & 15 -- 1301 & 27 & J0423-0120 & J0438+3004 & J0423-0120 \\
& 08/20/2022 & 47 & 15 -- 1301 & 27 & J0423-0120 & J0438+3004 & J0423-0120 \\
HD 163296 & 05/15/2022 & 43 & 15 -- 740 & 27 & J1924-2914 & J1742-1517 & J1924-2914 \\
& 05/15/2022 & 44 & 15 -- 740 & 27 & J1924-2914 & J1742-1517 & J1924-2914 \\
LkCa 15 & 01/06/2022 & 37 & 14 -- 976 & 29 & J0510+1800 & J0438+3004 & J0510+1800 \\
& 08/05/2022 & 45 & 15 -- 1301 & 29 & J0510+1800 & J0438+3004 & J0510+1800 \\
MWC 480 & 08/16/2022 & 42 & 15 -- 1301 & 35 & J0510+1800 & J0438+3004 & J0510+1800 \\
& 08/17/2022 & 43 & 15 -- 1301 & 35 & J0510+1800 & J0438+3004 & J0510+1800 \\
&  08/19/2022 & 44 & 15 -- 1301 & 35 & J0510+1800 & J0438+3004 & J0510+1800 \\
V4046 Sgr & 04/19/2022 & 49 & 14 -- 500 & 33 & J1924-2914 & J1826-3650 & J1924-2914 
\enddata
\end{deluxetable}

\begin{table}
    \centering
    \caption{Band 7 continuum observations.}
    \label{tab:cont}
    \begin{tabular}{lcc}
        \hline
        Source & Flux  & Beam (PA) \\
               & (mJy)  &    \\
        \hline
        AS 209 & 430 $\pm$ 43  & 0.47$\times$0.33 (63.8$^\circ$) \\
        DM Tau & 161 $\pm$ 21 & 0.38$\times$0.29 (-17.3$^\circ$)\\
        GM Aur & 545 $\pm$ 56 & 0.47$\times$0.26 (3.9$^\circ$)\\
        HD 163296 & 1536 $\pm$ 168 & 0.39$\times$0.31 (-51.2$^\circ$)\\
        LkCa 15 & 456 $\pm$ 46 & 0.38$\times$0.29 (-42.4$^\circ$)\\
        MWC 480 & 617 $\pm$ 62 & 0.30$\times$0.21 (15.8$^\circ$)\\
        V4046 Sgr & 776 $\pm$ 85 & 0.84$\times$0.53 (88.0$^\circ$)\\
        \hline
    \end{tabular}
    \footnotesize{Errors include the nominal 10$\%$ uncertainty from ALMA.}
\end{table}

\section{Disk-Integrated Spectra}\label{app:spec}

\begin{figure*}
    \centering
    \includegraphics[scale=0.25]{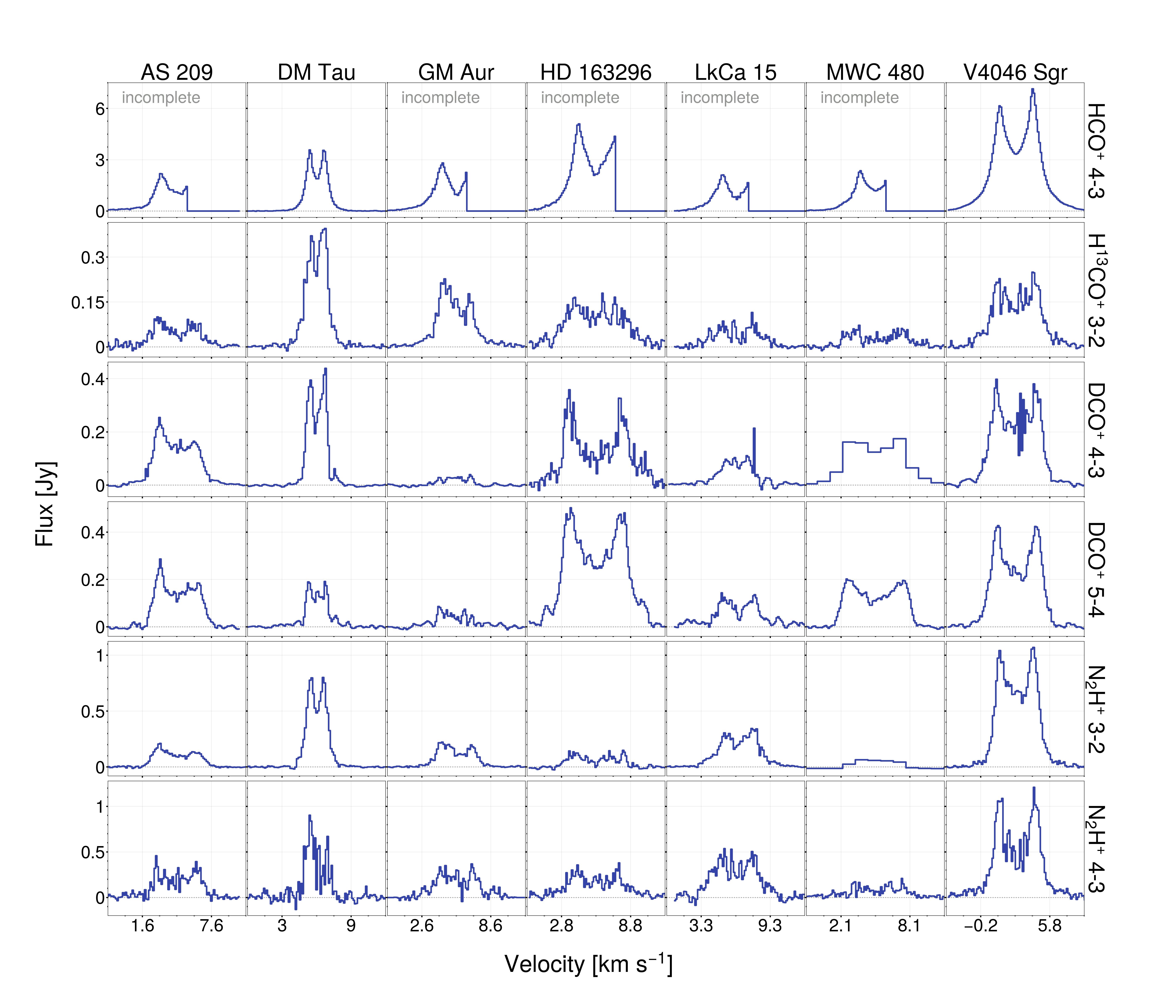}
        \caption{\normalsize{Disk-integrated spectra. We note the data cut-off issue which affects the \hcop\ data for AS 209, GM Aur, HD 163296, LkCa 15, and MWC 480 (2021.1.00138.S).}}
  \label{fig:spec}
\end{figure*}

Figure \ref{fig:spec} shows the disk-integrated spectra for all of the lines in our survey. The \hcop\ data issue can be seen clearly in the spectra, where some red-shifted velocities are cut off in AS 209, GM Aur, HD 163296, LkCa 15, and MWC 480. The complete \hcop\ spectral profiles for DM Tau and V4046 Sgr are fairly symmetric so we operate under the assumption of symmetry when retrieving integrated fluxes and radial profiles for the remaining sources. 

\section{Table of molecular ion substructure locations.}\label{app:substr}

A full list of radial molecular ion substructures can be found in Table \ref{tab:ion_substr}. We do not include estimates of feature widths or depths given the varying resolution of our survey data. Errors on the radial locations are taken to be a quarter of the beam size in au. 

\begin{table*}
\centering
\caption{Radial locations of molecular ion substructures.}
\label{tab:ion_substr}
\begin{minipage}[t]{0.48\linewidth}
    \centering
    \begin{tabular}{llcc}
\hline\hline
Source & Transition & Feature ID & Radius (au)\\
\hline
AS 209 & H$^{13}$CO$^{+}$ $3 - 2$ & B45 & 45 $\pm$ 15.4\\
& N$_{2}$H$^{+}$ $3 - 2$ & B70 & 70 $\pm$ 11.6\\
& N$_{2}$H$^{+}$ $4 - 3$ & B67 & 67 $\pm$ 15.2\\
& DCO$^{+}$ $4 - 3$ & B66 & 66 $\pm$ 16.3\\
& DCO$^{+}$ $5 - 4$ & B67 & 67 $\pm$ 16.0\\
\hline
DM Tau & H$^{13}$CO$^{+}$ $3 - 2$ & B18 & 18 $\pm$ 10.8\\
& N$_{2}$H$^{+}$ $3 - 2$ & B98 & 98 $\pm$ 7.8\\
& N$_{2}$H$^{+}$ $4 - 3$ & B92 & 92 $\pm$ 15.9\\
& DCO$^{+}$ $5 - 4$ & D112 & 112 $\pm$ 15.2\\
& & B228 & 228 $\pm$ 15.2\\
\hline
GM Aur & H$^{13}$CO$^{+}$ $3 - 2$ & B126 & 126 $\pm$ 16.5\\
& N$_{2}$H$^{+}$ $3 - 2$ & B84 & 84 $\pm$ 12.4\\
& N$_{2}$H$^{+}$ $4 - 3$ & B78 & 78 $\pm$ 37.0\\
& DCO$^{+}$ $5 - 4$ & D221 & 221 $\pm$ 25.7\\
& & B336 & 336 $\pm$ 25.7\\
\hline
LkCa 15 & H$^{13}$CO$^{+}$ $3 - 2$ & B50 & 50 $\pm$ 23.7\\
& N$_{2}$H$^{+}$ $3 - 2$ & B58 & 58 $\pm$ 11.4\\
& N$_{2}$H$^{+}$ $4 - 3$ & B58 & 58 $\pm$ 13.8\\
& DCO$^{+}$ $5 - 4$ & B54 & 54 $\pm$ 23.8\\
\hline
V4046 Sgr & H$^{13}$CO$^{+}$ $3 - 2$ & B48 & 48 $\pm$ 10.9\\
& N$_{2}$H$^{+}$ $3 - 2$ & B66 & 66 $\pm$ 21.6\\
& N$_{2}$H$^{+}$ $4 - 3$ & B57 & 57 $\pm$ 21.6\\
& DCO$^{+}$ $4 - 3$ & B61 & 61 $\pm$ 10.2\\
& DCO$^{+}$ $5 - 4$ & B68 & 68 $\pm$ 14.5\\
\hline
\end{tabular}%
\end{minipage}%
\hfill%
\begin{minipage}[t]{0.48\linewidth}
    \centering
\begin{tabular}{llcc}
\hline\hline
Source & Transition & Feature ID & Radius (au)\\
\hline
MWC 480 & HCO$^{+}$ $4 - 3$ & B160 & 160 $\pm$ 13.7\\
& & D232 & 232 $\pm$ 13.7\\
& & B267 & 267 $\pm$ 13.7\\
& H$^{13}$CO$^{+}$ $3 - 2$ & B52 & 52 $\pm$ 30.6\\
& N$_{2}$H$^{+}$ $3 - 2$ & B202 & 202 $\pm$ 46.2\\
& N$_{2}$H$^{+}$ $4 - 3$ & B102 & 102 $\pm$ 17.7\\
& DCO$^{+}$ $5 - 4$ & B76 & 76 $\pm$ 23.9\\
\hline
HD 163296 & HCO$^{+}$ $ 4 - 3$ & D90 & 90 $\pm$ 10.7\\
& & B125 & 125 $\pm$ 10.7\\
& H$^{13}$CO$^{+}$ $3 - 2$ & B97 & 97 $\pm$ 14.9\\
& & D235 & 235 $\pm$ 14.9\\
& & B310 & 310 $\pm$ 14.9\\
& N$_{2}$H$^{+}$ $3 - 2$ & B146 & 146 $\pm$ 13.8\\
& & B263 & 263 $\pm$ 13.8\\
& N$_{2}$H$^{+}$ $4 - 3$ & D53 & 53 $\pm$ 13.6\\
& & B86 & 86 $\pm$ 13.6\\
& & D202 & 202 $\pm$ 13.6\\
& & B236 & 236 $\pm$ 13.6\\
& DCO$^{+}$ $ 4 - 3$ & B49 & 49 $\pm$ 11.9\\
& & D85 & 85 $\pm$ 11.9\\
& & B110 & 110 $\pm$ 11.9\\
& DCO$^{+}$ $5 - 4$ & B49 & 49 $\pm$ 11.0\\
& & D78 & 78 $\pm$ 11.0\\
& & B110 & 110 $\pm$ 11.0\\
\hline
\end{tabular}
\end{minipage}
\end{table*}

\section{Ion flux correlations with source parameters}\label{app:fluxcorr}

\begin{figure*}
    \centering
    \includegraphics[width=\textwidth]{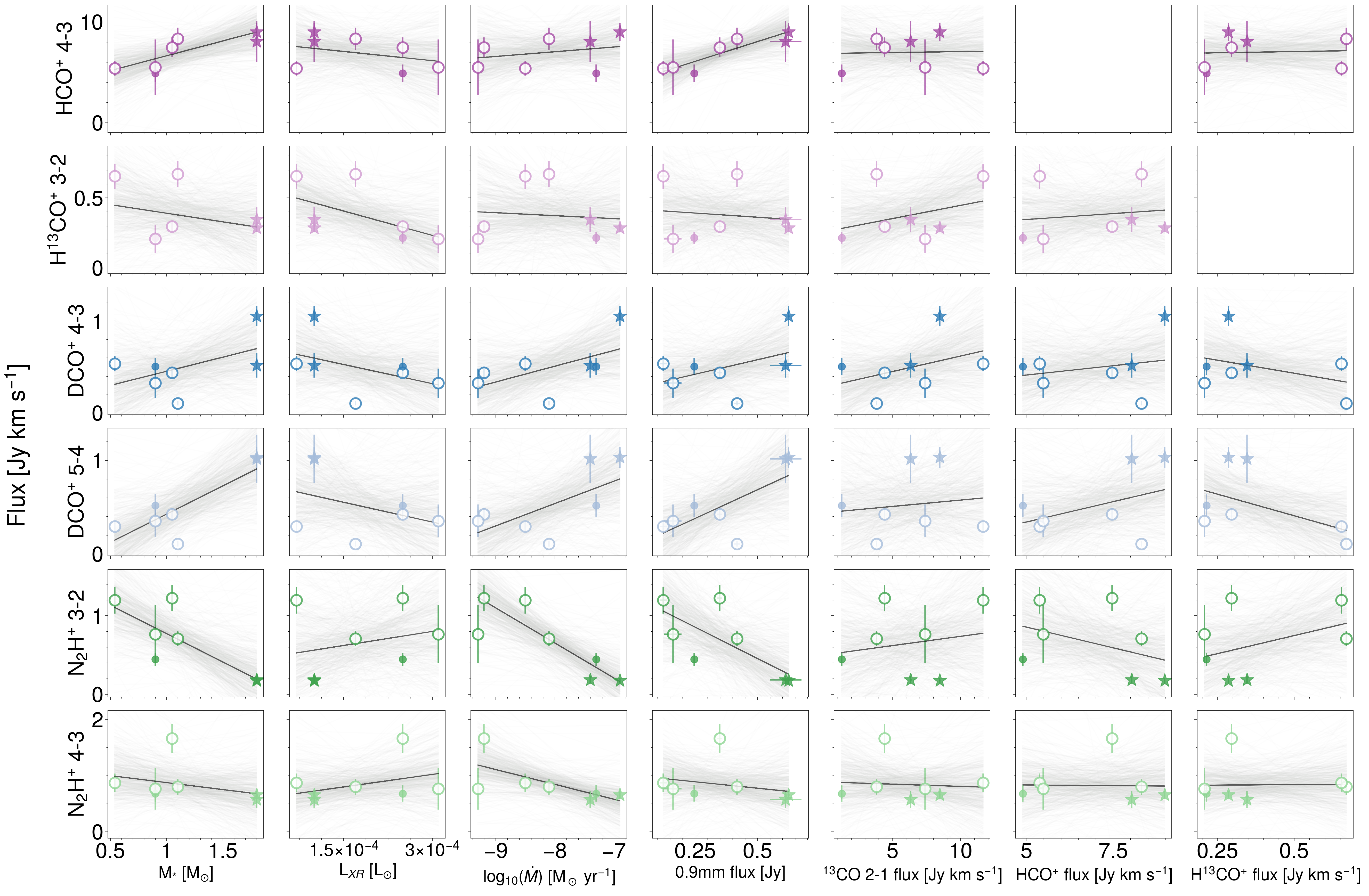}
        \caption{\normalsize{Line fluxes vs. source properties: stellar mass, stellar X-ray luminosity, mass accretion rate, 0.9 mm continuum flux, $^{13}$CO flux, \hcop\ flux, and \htcop\ flux. All fluxes are scaled to a common distance of 160 pc. Marker shapes indicate different source types: filled circles are continuous T-Tauris, open circles are transition disks, and stars are Herbigs. 0.9 mm continuum flux values were measured from our Band 7 observations and are reported in Table \ref{tab:cont}. $^{13}$CO $J = 2 - 1$ flux values were retrieved from \citet{oberg_maps_2021} for AS 209, GM Aur, HD 163296, and MWC 480, \citet{long_22} for DM Tau and V4046 Sgr, and \citet{leemker_2022} for LkCa 15. The black line shows the median fit from \texttt{linmix} and the grey lines show samples from the posterior distribution.}}
  \label{fig:fluxcorr}
\end{figure*}

Figure \ref{fig:fluxcorr} shows the relationships between the measured molecular ion fluxes and different source properties. 

\end{document}